\documentclass[aps,prd,twocolumn,floatfix,superscriptaddress,showpacs,showkeys,amsmath,amssymb]{revtex4} 
\usepackage{graphicx}
\usepackage{dcolumn}
\usepackage{bm}
\usepackage[dvips]{epsfig}
\usepackage{epsfig}
\usepackage{graphics}
\usepackage{subfigure}
\usepackage{mathrsfs}
\usepackage{amssymb,latexsym}

\newcommand{\seq}{\begin{subequations}}
\newcommand{\sen}{\end{subequations}}
\newcommand{\eq}{\begin{eqnarray}}
\newcommand{\en}{\end{eqnarray}}
\newcommand{\ra}{\rangle} 
 
\def\shiftleft#1{#1\llap{#1\hskip 0.04em}}
\def\shiftdown#1{#1\llap{\lower.04ex\hbox{#1}}}
\def\thick#1{\shiftdown{\shiftleft{#1}}}
\def\b#1{\thick{\hbox{$#1$}}}
\begin{document}
\title{Electromagnetic structure of the nucleon and the Roper resonance\\ 
in a light-front quark approach} 

\author{Igor T. Obukhovsky} 
\affiliation{Institute of Nuclear Physics, Moscow
State University,119991 Moscow, Russia}
\author{Amand Faessler} 
\affiliation{
Institut f\"ur Theoretische Physik, Universit\"at T\"ubingen,\\
Kepler Center for Astro and Particle Physics,\\
Auf der Morgenstelle 14, D-72076, T\"ubingen, Germany}
\author{Thomas Gutsche}
\affiliation{
Institut f\"ur Theoretische Physik, Universit\"at T\"ubingen,\\
Kepler Center for Astro and Particle Physics,\\
Auf der Morgenstelle 14, D-72076, T\"ubingen, Germany}
\author{Valery E. Lyubovitskij} 
\affiliation{
Institut f\"ur Theoretische Physik, Universit\"at T\"ubingen,\\
Kepler Center for Astro and Particle Physics,\\
Auf der Morgenstelle 14, D-72076, T\"ubingen, Germany}
\affiliation{Department of Physics, Tomsk State University,
634050 Tomsk, Russia}

\date{\today}

\begin{abstract}
A relativistic light-front quark model is used to describe both the 
elastic nucleon and nucleon-Roper transition form factors 
in a large $Q^2$ range, up to 35 GeV$^2$ for the
elastic and up to 12 GeV$^2$ for the resonance case. Relativistic
three-quark configurations satisfying the Pauli exclusion principle on the 
light-front are used for the derivation of the current matrix elements.
The Roper resonance is considered as a mixed state of a three-quark core 
configuration and a molecular $N+\sigma$ hadron component. Based on 
this ansatz we obtain a realistic description of both processes, elastic
and inelastic, and show that existing experimental data are indicative of a
composite structure of the Roper resonance. 
\end{abstract}

\pacs{12.39.Ki, 13.40.Gp, 13.40.Hq, 14.20.Gk}

\keywords{Roper resonance, quark model, hadron molecules, 
strong and electromagnetic form factors} 

\maketitle

\section{Introduction.}\label{s1}
 
The last decade has been marked by significant progress in the experimental 
study of low-lying baryonic resonances (the radial/orbital nucleon 
excitations with $J^P=\frac{1}{2}^{\pm}$, $\frac{3}{2}^{\pm}$). 
Specifically new insights have been obtained
in $\pi$~\cite{aznauryan09} and $2\pi$~\cite{mokeev09} electroproduction on 
the proton with the polarized electron beam at  JLab (CLAS Collaboration) 
followed by a combined analysis of pion- and photoinduced reactions made by 
CB-ELSA and the A2-TAPS collaborations~\cite{sarantsev08}. Electro- and 
photoproduction of these resonances is recognized as an important tool
which allows to study the relevant degrees of freedom, wave functions and
interactions between constituents and the transition to perturbative 
quantum chromodynamics (pQCD).

The structure issue of the lowest-lying nucleon re\-so\-nan\-ce $N(1440)$ with
$J^P = \frac{1}{2}^+$  (the Roper resonance $P_{11}$ or simply $R$) is a 
longstanding problem of hadron physics. One indication that the inner 
structure of the Roper is possibly more complicated than the structure of the 
other lightest baryons was first obtained in the framework of the 
constituent quark model (CQM). It was found that the observed mass 
of the Roper resonance is much too low and the decay width is too large 
when compared to the predicted values of the CQM.
The simplest description of the Roper consists of the three-quark $(3q)$ 
configuration $sp^2[3]_X$, i.e. the first ($2S$) radial excitation of the 
nucleon ground state $s^3[3]_X$, but it fails to explain either the large 
decay width $\Gamma_R\simeq$ 300 MeV or the branching ratios for the $\pi N$ 
(55\%--75\%) and $\sigma N$ (5\%--20\%) 
decay channels~\cite{pdg10,sarantsev08}. 

Evaluation of these values in the framework of the CQM is often based on 
the elementary emission model with single-particle quark-meson (or 
quark-gamma) couplings $qq\pi$, $qq\sigma$, $qq\gamma$, etc. The calculation 
of decay widths (or of the  electroproduction cross section at
small virtuality of the photon with $Q^2\simeq\,$0) results in anomalous 
small values. These underestimates for the decay matrix elements
can especially be traced to the strict
requirement of orthogonality for the ground ($0S$) and excited state ($2S$) 
radial wave functions of the $N$- and $R$ states belonging 
to quark configurations with the same spin-isospin ($S=1/2$, $T=1/2$) 
and symmetry ($[3]_{ST}[3]_X$) quantum numbers. To overcome this discrepancy
it is suggested that either the Roper is not an ordinary $3q$ state or the 
"true" transition operators have a more complicated form than the 
single-particle operators used in the CQM calculations. 

The elementary theory of strong interactions QCD provides a framework,
which is directly usable only at high momentum transfers. Nevertheless, the 
discussed data~\cite{aznauryan09,mokeev09,sarantsev08} span the range from 
soft to hard momentum transfers 0$\,\le Q^2\lesssim\,$4 -- 5 GeV$^2$ (up to 
$\sim$ 12 GeV$^2$ for the JLab upgrade). A major challenge for theory is 
that a quantitative description of the transition amplitudes must also include 
soft nonperturbative contributions. For the soft region there are important 
results from lattice QCD with ``unquenched'' $\bar qq$ degrees of 
freedom~\cite{bowman04,engel11} but the present computer capabilities do not 
allow us yet to extract all the hadron properties in a systematic way.

Other approaches that are directly connected to QCD are either based on
light cone sum rules~\cite{braun09} (in reality they can be used at 
$Q^2\gtrsim m^2_{N^*}$) or on Dyson-Schwinger equations 
(DSEs)~\cite{roberts94,roberts11,roberts11a}. A DSE study~\cite{roberts11a} 
produces a radial excitation of the nucleon in the quark-diquark basis at 
$\sim$1.82 GeV. Pion electroproduction amplitudes in the resonance 
region $W\simeq m_R$ are successfully analyzed in terms of the dynamical 
coupled channel model~\cite{matsuyama07,suzuki09,suzuki10}, which is used at 
the Excited Baryon Analysis Center at JLab. Combining both methods,  
Refs.~\cite{roberts11,suzuki10} demonstrate that the Roper resonance is 
indeed the first radial excitation of the proton, but the Roper ``obscures its 
dressed-quark core with a dense cloud of pions and other 
mesons''\cite{roberts11}.

The $\sigma$ meson along with the uncorrelated pion cloud possibly
play a key role in the inner structure of the Roper. This mechanism 
was proposed in Ref.~\cite{hanhart00} where the authors showed that 
in $\pi N$ scattering
the intermediate $\sigma N$ state defines the Roper resonance pole.
Thus there is no need for some special quark configuration of the type 
$sp^2[3]_X$ to describe the Roper resonance contribution to 
the physical processes.
 
It is clear that the nature of the low-lying baryonic resonances
is still an unresolved issue and in this respect the study of the 
$Q^2$-behavior of their electroproduction amplitude is of much current 
interest. Since direct QCD calculations are difficult in 
the low-energy regime several models for the electroexcitation
of the Roper resonance were proposed during 
the last three decades~\cite{gavella80,li90,close90,capstick95,%
cardarelli97,aznauryan07,cano98,riska06}
(see reviews~\cite{burkert04,aznauryan12a} for details). Now model predictions 
can be compared to the new high-quality photo- and electroproduction 
data~\cite{aznauryan09,mokeev09,sarantsev08}.
Updated versions~\cite{capstick07,aznauryan12,ramalho10,golli09} of the most
realistic models were used to give a good description of the data at 
intermediate values of 1.5$\,\lesssim Q^2\lesssim\,$4~GeV$^2$.
However, in the soft region, i.e. at low values of $Q^2$ 
(0$\,\leq Q^2\lesssim\,$1 -- 1.5 GeV$^2$), the data qualitatively differ from 
theoretical predictions made in the framework of quark models without meson 
cloud.

Recently the electromagnetic nucleon-Roper transition 
has been studied in the framework of anti-de Sitter 
AdS/QCD~\cite{deTeramond:2011qp,Gutsche:2012wb}. 
In particular, in Ref.~\cite{deTeramond:2011qp} the Dirac form factor for the 
electromagnetic nucleon-Roper transition has been calculated in 
light-front holographic QCD. In Ref.~\cite{Gutsche:2012wb} 
the Roper electroproduction was considered in a soft-wall 
AdS/QCD model~\cite{Approach,Approach2,Gutsche:2012bp} with 
inclusion of the leading three-quark ($3q$) state and higher Fock components. 

As a result there are essentially three comprehensive theoretical approaches 
to the Roper electroproduction on the market. One of them (the coupled channel 
model of the meson cloud~\cite{mokeev09,matsuyama07,suzuki09,kamano10}) is  
successful in the soft region 0$\,\leq Q^2\lesssim\,$1 GeV$^2$ 
and, the other one, the LF three-quark 
model~\cite{capstick07,cardarelli97,aznauryan07,aznauryan12} or the covariant 
quark spectator model~\cite{ramalho10}) is compatible with data in the hard 
region $\,Q^2\gtrsim\,$$m_N^2$-$2m_N^2$. The third approach is based on 
a novel method to hadronic structure --- 
AdS/QCD~\cite{deTeramond:2011qp,Gutsche:2012wb}.   

In our recent work~\cite{obukhovsky11} we obtained a quantitative 
description of the Roper electroproduction helicity amplitudes in the 
region 0$\,\le Q^2\lesssim\,$2 GeV$^2$ where we started with the following 
model principles: 

\noindent
({\it i}) ''Unquenching'' of the constituent quark model, i.e. taking into 
account the $q\bar q$ pair effects (e.g., see the discussion in 
Ref.~\cite{BRAG07}) in the soft-$Q^2$ region. This leads to a nonlocal 
$qq\gamma$ coupling depending on the inner momentum of the $q\bar q$
wave function of the intermediate vector meson [the vector meson 
dominance (VMD) is implied]. 

\noindent
({\it ii}) Smooth transition from the ``soft'' nonlocal  electromagnetic 
coupling to the ``hard'' one with growing momentum transfer $Q^2$. 
In the hard region the nonlocal $qq\gamma$ coupling reduces to the 
standard $Q^2$-dependent quark form factor characteristic of the VMD model.

\noindent
({\it iii}) The hadron-molecular $N+\sigma$ state is considered as a possible 
component of the Roper wave function along with the radial excitation of the 
three-quark configuration.

Although nonrelativistic quark configurations were used, a realistic 
description of the $Q^2$ dependence of transition amplitudes was obtained in 
a large interval of momentum transfers $0 \le Q^2 \lesssim  1.5 -- 2$ GeV$^2$. 
Given the quality of the suggested model it reinforces the statement 
that symmetry principles (e.g. the Pauli principle for quark systems 
including the $3q+q\bar q$ component, the VMD in the 
electromagnetic coupling, etc.) play the decisive role in the description of 
the electroexcitation of low-lying resonances.

Starting from the results of Ref.~\cite{obukhovsky11} we developed 
a relativistic version of the suggested electroexcitation mechanism. 
Wave functions of baryons are set up in a LF constituent 
quark model based on the relativistic Hamiltonian dynamics which 
was first formulated 
by Berestetskii and Terent'ev~\cite{berestetskii76} and applied to various 
hadronic processes in Refs.~\cite{capstick95,capstick07,bakker79,aznauryan82,%
aznauryan12,keister88,cardarelli97,Schlumpf95,Schlumpf92}. 

The paper is structured as follows. First, in Sec.~\ref{s2}, we briefly 
discuss the LF formalism relevant for the electroproduction processes.
In Sec.~\ref{s3} we fit the parameters 
of the model to the elastic $e-N$ data (up to $Q^2\approx$ 35 GeV$^2$) 
including the limit $Q^2\to$ 0 (magnetic moments). 
The quality of fit in describing the nucleon form 
factors is a test of our version of the LF approach to the nucleon 
electromagnetic coupling. We further use this model in Sec. III for the
description of the quark core contribution to the electroproduction  
amplitudes. 
In the framework of our model for the Roper resonance considered 
as a composite 
state~\cite{obukhovsky11} we calculate the helicity amplitudes $A_{1/2}$ 
(transverse) and $S_{1/2}$ (longitudinal) for the Roper resonance 
electroproduction on the nucleon.  The results obtained are compared to the 
recent CLAS data (up to $Q^2\approx$ 4 -- 5 GeV$^2$). Predictions for 
higher values of $Q^2$ (up to 12 GeV$^2$ for the JLab upgrade) are 
also discussed. Finally, in Sec.~\ref{s4}, we summarize our results.

\section{Definition of form factors in terms of three-quark configurations on 
the light-front}\label{s2}

The light-front approach~\cite{berestetskii76,bakker79,aznauryan82,keister88} 
to elastic and inelastic nucleon form factors was
used in many works in the last three 
decades~\cite{capstick95,cardarelli97,aznauryan07,capstick07,%
BRAG07,aznauryan12,aznauryan12a,keister88,Schlumpf95,Schlumpf92}. 
We follow Refs.~\cite{capstick95,cardarelli97,keister88,Schlumpf95,Schlumpf92} 
where the method was described 
in many details. Here we only accentuate some aspects which are not covered
in the literature but which are essential for us in the study of the nucleon 
form factors. In addition in Sect.~\ref{s23} we cite several well known 
formulas from Ref.~\cite{keister88} to allow an easier reading of our 
manuscript.

\subsection{Melosh rotation in the $3q$ system}\label{s21}

First of all it should be noted that the Melosh rotation
\begin{equation}
R_M^{(i)}(x_i,{\b k}_{\bot i},{\cal M}_0)=
\frac{m_i+x_i{\cal M}_0-i\bm{\sigma}\cdot[\bm{\hat z}\times\bm{k}_{\bot i}]}
{\sqrt{(m_i+x_i{\cal M}_0)^2+\bm{k}_{\bot i}^2}}
\label{mel}
\end{equation}
(we use standard kinematical parameters $m_i$,$x_i$ and $\b{k}_{\bot i}$
of the $i$th quark, which are defined below) 
is not a trivial operation in the case of a three-fermion system which should
satisfy the Pauli exclusion principle. Sometimes the product of three 
independent rotations 
\begin{equation}
{\cal R}_M=\prod_{i=1}^3R_M^{(i)}
\label{3mel}
\end{equation}
produces a change in the type of initial permutational symmetry (the Young 
scheme) of the $3q$ wave function.

 If one suggests that the wave function is defined by a certain LF dynamics 
such a function should satisfy the Pauli exclusion principle. In reality we 
start from the canonical ($c$) quark spin wave function defined in the rest 
frame, where the fully symmetric spin-isospin ($ST$) state of three quarks
(the Young scheme $[3]_{\scriptscriptstyle ST}$) has the following (allowed
by the Pauli principle) form:
\begin{multline}
|[3]_{\scriptscriptstyle ST},\mu^\prime,t{\rangle_c}=\sqrt{\frac{1}{2}}
|[21]_{\scriptscriptstyle S}y^{(1)}_{\scriptscriptstyle S},
\mu^\prime{\rangle_c}
|[21]_{\scriptscriptstyle T}y^{(1)}_{\scriptscriptstyle T},t\rangle\\
+\sqrt{\frac{1}{2}}
|[21]_{\scriptscriptstyle S}y^{(2)}_{\scriptscriptstyle S},
\mu^\prime{\rangle_c}
|[21]_{\scriptscriptstyle T}y^{(2)}_{\scriptscriptstyle T},t\rangle
\label{cpau}
\end{multline}
[for S=1/2 (or $[21]_{\scriptscriptstyle S}$) and T=1/2 (or
$[21]_{\scriptscriptstyle T}$)]. Here we use the Yamanuchi symbol $y^{(i)}$ 
(see Ref. ~\cite{hamermesh64} for details)
for a compact representation of the sequence of spin
couplings in the $3q$ states $|s_1s_2(S_{12})s_3:\!S,\mu^\prime{\rangle_c}$,
\begin{eqnarray}
|[21]_{\scriptscriptstyle S}y^{(1)}_{\scriptscriptstyle S},
\mu^\prime{\rangle_c}&=&
|\frac{1}{2}\frac{1}{2}(1)\frac{1}{2}:\frac{1}{2},\mu^\prime{\rangle_c}
\nonumber\\
|[21]_{\scriptscriptstyle S}y^{(2)}_{\scriptscriptstyle S},
\mu^\prime{\rangle_c}&=&
|\frac{1}{2}\frac{1}{2}(0)\frac{1}{2}:\frac{1}{2},\mu^\prime{\rangle_c},
\label{yam}
\end{eqnarray}
(the same notations are used for isospin states).

As usual the $z$ axis is taken as the quantization axis both for the canonical 
($c$) and the front form ($f$) spins. Here and further on, the symbols 
$\mu_i^\prime,\mu_{12}^\prime,\mu^\prime$ and 
$\mu_i^{\prime\prime},\mu_{12}^{\prime\prime},\mu^{\prime\prime}$ denote 
the canonical spin projections on the $z$ axis in initial and final states, 
respectively, while $\mu_i,\mu_{12},\mu$ and $\bar\mu_i,\bar\mu_{12},\bar\mu$ 
are the front form spin projections. Under the Melosh rotations (different for
each quark with label $i$)
\begin{equation}
R_M^{(i)}|\frac{1}{2},\mu_i^\prime{\rangle_c}=\sum_{\mu_i} 
D^{(\frac{1}{2})}_{\mu_i,\mu_i^\prime}(\theta^{(i)}_M)
|\frac{1}{2},\mu_i{\rangle_f}
\label{imel}
\end{equation}
the canonical spin basis functions (\ref{yam}) are transformed
into the front form
\begin{multline}
{\cal R}_M|\frac{1}{2}\frac{1}{2}(S^\prime_{12})\frac{1}{2}:
S^\prime,\mu^\prime
\rangle_{c}=\sum_{S=1/2,3/2}\,\, \sum_{S_{12}=0,1}\\
\times\sum_\mu C^{SS^\prime}_{S_{12},S^\prime_{12}}(\mu,\mu^\prime)
|\frac{1}{2}\frac{1}{2}(S_{12})\frac{1}{2}:S,\mu\rangle_{\!f}\,.
\label{fc}
\end{multline}
The coefficients $C^{SS^\prime}_{S_{12},S^\prime_{12}}(\mu,\mu^\prime)$ 
are the matrix 
elements of the triple product (\ref{3mel}) of Melosh matrices (\ref{mel}) 
between the spin basis states in the $3q$ system (\ref{yam}). 
Explicit expressions for the coefficients 
$C^{SS^\prime}_{S_{12},S^\prime_{12}}(\mu,\mu^\prime)$ are given in 
the Appendix. The coefficients
$C^{SS^\prime}_{S_{12},S^\prime_{12}}$ depend on the relative momenta of quarks
\begin{eqnarray}
\bm{\lambda}_\bot&=&\frac{x_2\bm{k}_{\bot 1}-x_1\bm{k}_{\bot 2}}{x_1+x_2},\quad
m=m_1=m_2=m_3,\nonumber\\
\bm{\Lambda}_\bot&=&\frac{x_3(\bm{k}_{\bot 1}+\bm{k}_{\bot 2})-
(x_1+x_2)\bm{k}_{\bot 3}}{x_1+x_2+x_3}=-\bm{k}_{\bot 3}
\label{lam}
\end{eqnarray}
and on the $z^+$ components of the quark momenta, 
$x_i=p_i^+/P^+$$=k_i^+/{\cal M}_0$ with
\begin{equation}
x_1=\xi\eta,\quad x_2=\eta(1-\xi),\quad x_3=1-\eta \,,
\label{xi}
\end{equation}
where $\{x_i,\bm{k}_{\bot i}\}$ is the LF momentum of the $i$th 
quark defined through the Lorentz transformation of the $i$th 
quark momentum to the rest frame of the $3q$ system.

Note that the relative momentum $\bm{\lambda}$ is odd with respect to the 
permutation $P_{12}$ of the first and second quark while momentum 
$\bm{\Lambda}$ is even
\begin{equation}
P_{12}\bm{\lambda}_\bot=-\bm{\lambda}_\bot,\quad
P_{12}\bm{\Lambda}_\bot=\bm{\Lambda}_\bot 
\label{odd}
\end{equation}
with $P_{12}\xi=1-\xi$, $P_{12}\eta=\eta$.
Thus the permutation symmetry of the resulting LF state (\ref{fc})
is not trivial and should be considered in detail.

In our case we have $S^{\,\prime}=\frac{1}{2}$ (the canonical nucleon spin), 
but after Melosh rotations of the quark spins this value transforms into 
two different values, $S=\frac{1}{2}$ and $\frac{3}{2}$, which characterize 
the two different components of the nucleon wave function in the front form. 
For example, if we start from the canonical spin state 
$|\frac{1}{2}\frac{1}{2}(0)\frac{1}{2}:\frac{1}{2},\mu^\prime{\rangle_c}$ 
[which possesses the fixed permutation symmetry
$[21]_{\scriptscriptstyle S}y^{(2)}_{\scriptscriptstyle S}$] we obtain the 
front-spin states with other values of $S_{12}$ or $S$, e.g. $S_{12}=1$ and 
$S=\frac{3}{2}$ [i.e. the states 
$|[21]_{\scriptscriptstyle S}y^{(1)}_{\scriptscriptstyle S}, 
\mu^\prime{\rangle_f}$ and $|[3]_{\scriptscriptstyle S}, 
\mu^\prime{\rangle_f}$] which supposedly violate the initial permutation 
symmetry of the $3q$ system. 

However, on the light front this does not lead to a violation of the Pauli 
exclusion principle. The reason for this is that the Melosh rotation of the 
spin state $|[21]_{\scriptscriptstyle S}y^{(i)}_{\scriptscriptstyle S},
\mu{\rangle_c}$ turns into a superposition of combined spin-orbital ($SP$) 
states $|[21]_{\scriptscriptstyle PS}y^{(j)}_{\scriptscriptstyle PS},
\mu{\rangle_f} (j=1,2)$ realized in the product space of spin $(S)$ and 
momentum $(P)$.
The momentum-dependent factors of the coefficients $C^{SS^\prime}_{00}$ and 
$C^{SS^\prime}_{11}$ are even with respect to the permutation $P_{12}$ 
while the coefficients $C^{SS^\prime}_{10}$ and $C^{SS^\prime}_{01}$ are odd.
Hence, e.g. in the case of $C^{\frac{1}{2}\frac{1}{2}}_{10}$, the term 
$\sum_\mu C^{\frac{1}{2}\frac{1}{2}}_{10}(\mu,\mu^\prime)|
[21]_{\scriptscriptstyle S}
y^{(2)}_{\scriptscriptstyle S}, \mu^\prime{\rangle_f}$ in Eq.~(\ref{fc}) 
has the 
same value ($y_{\scriptscriptstyle PS}^{(2)}$) of the Yamanuchi symbol in the 
(front) $SP$ space as the initial value $y_{\scriptscriptstyle S}^{(2)}$ 
in the (canonical) $S$ space.

Using the coefficients $C^{\frac{1}{2}\frac{1}{2}}_{00}$, 
$C^{\frac{1}{2}\frac{1}{2}}_{11}$, $C^{\frac{3}{2}\frac{1}{2}}_{11}$,
$C^{\frac{1}{2}\frac{1}{2}}_{10}$, $C^{\frac{3}{2}\frac{1}{2}}_{10}$ and
$C^{\frac{1}{2}\frac{1}{2}}_{01}$
in Eq.~(\ref{fc}) we obtain the correct basis vectors in the product space
(the LF spin $S$ and the LF momentum $P$) for the irreducible 
representation (IR) $[21]_{PS}$ of the permutation group $S_3$
\begin{multline}
{\cal R}_M
|[21]_{\scriptscriptstyle S}y^{(1)}_{\scriptscriptstyle S},
\mu^\prime{\rangle_c}=\\
|[21]_{\scriptscriptstyle PS}y^{(1)}_{\scriptscriptstyle PS},
\mu^\prime{\rangle}_f=\sum_{\mu}\left[C^{\frac{1}{2}
\frac{1}{2}}_{01}(\mu,\mu^\prime)
|\frac{1}{2}\frac{1}{2}(0)\frac{1}{2}\!:\!\frac{1}{2},\mu{\rangle_f}+\right.\\
\left.C^{\frac{1}{2}\frac{1}{2}}_{11}(\mu,\mu^\prime)
|\frac{1}{2}\frac{1}{2}(1)\frac{1}{2}\!:\!\frac{1}{2},\mu{\rangle_f}\!
+C^{\frac{3}{2}\frac{1}{2}}_{11}(\mu,\mu^\prime)
|\frac{1}{2}\frac{1}{2}(1)\frac{1}{2}\!:\!\frac{3}{2},\mu{\rangle_f}\!\!
\right]
\label{s3s1}
\end{multline}
and
\begin{multline}
{\cal R}_M
|[21]_{\scriptscriptstyle S}y^{(2)}_{\scriptscriptstyle S},
\mu^\prime{\rangle_c}=\\
|[21]_{\scriptscriptstyle PS}y^{(2)}_{\scriptscriptstyle PS},
\mu^\prime{\rangle}_f=\sum_{\mu}\left[C^{\frac{1}{2}
\frac{1}{2}}_{00}(\mu,\mu^\prime)
|\frac{1}{2}\frac{1}{2}(0)\frac{1}{2}\!:\!\frac{1}{2},\mu{\rangle_f}+\right.\\
\left.C^{\frac{1}{2}\frac{1}{2}}_{10}(\mu,\mu^\prime)
|\frac{1}{2}\frac{1}{2}(1)\frac{1}{2}\!:\!\frac{1}{2},\mu{\rangle_f}\!
+C^{\frac{3}{2}\frac{1}{2}}_{10}(\mu,\mu^\prime)
|\frac{1}{2}\frac{1}{2}(1)\frac{1}{2}\!:\!\frac{3}{2},\mu{\rangle_f}\!\!
\right]
\label{s3s2}
\end{multline}

The result of the Melosh rotation of the basis state (\ref{cpau}) can be 
written as a symmetric $SU(6)\times O(3)$ basis state $[3]_{PST}$ which 
satisfies the Pauli exclusion principle for the LF states:
\begin{multline}
{\cal R}_M|[3]_{ST},\mu^\prime t{\rangle_c}=\\ 
|[3]_{PST},\mu^\prime t{\rangle}_f
=\sqrt{\frac{1}{2}}|[21]_{\scriptscriptstyle PS}
y^{(1)}_{\scriptscriptstyle PS},\mu^\prime{\rangle_f}
|[21]_{\scriptscriptstyle T}y^{(1)}_{\scriptscriptstyle T},t\rangle\\
+\sqrt{\frac{1}{2}}|[21]_{\scriptscriptstyle PS}
y^{(2)}_{\scriptscriptstyle PS},\mu^\prime{\rangle_f}
|[21]_{\scriptscriptstyle T}y^{(2)}_{\scriptscriptstyle T},t\rangle \,.
\label{pst3}
\end{multline}

Here we consider the Clebsch-Gordon combinations of quark LF spins
\begin{multline}
|\frac{1}{2}\frac{1}{2}(S_{12})\frac{1}{2}:\frac{1}{2},\mu{\rangle_f}=
\sum_{\mu_{12}\mu_3}\sum_{\mu_1\mu_2}
(\frac{1}{2}\mu_1\frac{1}{2}\mu_2|S_{12}\mu_{12})\\
\times(S_{12}\mu_{12}\frac{1}{2}\mu_3|\frac{1}{2}\mu)|\frac{1}{2}
\mu_1{\rangle_f}
|\frac{1}{2}\mu_2{\rangle_f}|\frac{1}{2}\mu_3{\rangle_f} \, 
\label{fkg}
\end{multline}
used in the right-hand sides of Eqs. (\ref{s3s1}) and (\ref{s3s2}). They are
rather considered as basis vectors 
$|[21]_{S_{3}}y_{S_{3}}^{(i)},\mu{\rangle}$ of the IR $[21]_{S_{3}}$ of the 
symmetric group $S_3$ than the IR of the rotation group $O(3)$ [or the spin 
group $SU(2)_S$] which is not a kinematical subgroup for the LF dynamics. 
Fortunately, 
for three-particle systems the Clebsch-Gordon coefficients of the symmetric 
group $S_3$ are the same as the Clebsch-Gordon coefficients of the rotation 
group $O(3)$ [or the spin group $SU(2)$]~\cite{hamermesh64}.

\subsection{Nucleon and Roper resonance wave functions}\label{s22}

The spin-isospin part of the nucleon wave function is defined by 
Eq.~(\ref{cpau}) as a basis vector of the IR \underline{56} 
($J^P=\frac{1}{2}^+$) of the $SU(6)$ group. As a result of the Melosh rotation
we obtained in Eq.~(\ref{pst3}) a relativistic representation of this state
which depends on the light-front spin states defined in Eq.~(\ref{fkg}).

 The full wave function also possesses a scalar
factor $\Phi_S({\cal M}_0)$ --- the analog of the radial part of 
nonrelativistic wave function. To preserve relativistic invariance the LF wave 
function $\Phi_S$ should depend on the invariant mass of the system of 
initial quarks:
\begin{equation}
{\cal M}_0^2\equiv\sum_{i=1}^3\frac{m_i^2+\bm{k}_{\bot i}^2}{x_i}=
\frac{m^2+\bm{\lambda}_\bot^2}{\eta\xi(1-\xi)}+
\frac{\eta m^2+\bm{\Lambda}_\bot^2}{\eta(1-\eta)}
\label{M0}
\end{equation}
The mass ${\cal M}_0$ only depends on the square of relative momenta 
of quarks (\ref{lam}) and on the $z^+$ components (\ref{xi}).

Note that in a special Breit frame where the momentum $q^\mu$ 
transferred to the nucleon only has the transverse component 
$\bm{q}_\bot$ (for definiteness the momentum $\bm{q}_\bot$ is directed along 
the $x$ axis),
\begin{equation}
q^\mu=\{0,\bm{q}_\bot,0\},\,\,Q^2=-q^2=q^2_\bot,\,\, q_\bot=|\bm{q}_\bot|
\label{q}
\end{equation}
the quark relative momentum $\bm{\lambda}_\bot$ is not changed, while 
the values of $\bm{\Lambda}_\bot$ and ${\cal M}_0$ become 
$\bm{\Lambda}^\prime_\bot$ and ${\cal M}^\prime_0$, respectively, with
\begin{equation}
\bm{\Lambda}^\prime_\bot=\bm{\Lambda}_\bot-\eta\,\bm{q}_\bot,\quad
{{\cal M}^\prime_0}^2={{\cal M}_0}^2+\frac{\eta\,\bm{q}^2_\bot-2\bm{q}_\bot\!
\cdot\!\bm{\Lambda}_\bot}{1-\eta}\,.
\label{lamp}
\end{equation}

Using Eqs. (\ref{s3s1}) and (\ref{s3s2}) we can write the nucleon wave 
function in the form 
\begin{multline}
|N_{1/2^+}({\cal M}_0);\mu^\prime,t{\rangle}_f\\
=\Phi_S({\cal M}_0)\left[\sqrt{\frac{1}{2}}|[21]_{\scriptscriptstyle PS}
y^{(1)}_{\scriptscriptstyle PS},\mu^\prime{\rangle_f}\,
|[21]_{\scriptscriptstyle T}y^{(1)}_{\scriptscriptstyle T},t{\rangle}
\right.\\
\left.+\sqrt{\frac{1}{2}}|[21]_{\scriptscriptstyle PS}
y^{(2)}_{\scriptscriptstyle PS},
\mu^\prime{\rangle_f}\,
|[21]_{\scriptscriptstyle T}y^{(2)}_{\scriptscriptstyle T},t{\rangle}\right]\,.
\label{s3st}
\end{multline}
The Roper resonance wave function 
$|N^*_{1/2^+},{\cal M}_0;\mu^{\prime},t^{\prime}{\rangle}$ has the same form, 
but now the function $\Phi_S({\cal M}_0)$ corresponds to a 
radial excitation of the nucleon. We further denote the nucleon function 
as $\Phi_{0S}({\cal M}_0)$ and use the notation $\Phi_{2S}({\cal M}_0)$ for
the Roper resonance. 
In analogy to the harmonic oscillator model we define the radially excited 
wave function in the form
\begin{equation}
\Phi_{2S}={\cal N}_{2S}\left(1-c_R\frac{{\cal M}_0^2}{\beta^2}\right)
\Phi_{0S}\,,
\label{phi02s}
\end{equation}
where $\beta$ is the scale parameter.  
Here the coefficient $c_R$ can be determined by the orthogonality condition
\begin{equation}
{\langle}N^*_{1/2^+};\mu^\prime,t|N_{1/2^+};\mu^\prime,t{\rangle}=0 \,.
\label{cr}
\end{equation}
In the Breit frame (\ref{q}) the initial nucleon has momentum 
$-\frac{\bm{q}_\bot}{2}$ and the wave function (\ref{s3st}) is denoted
by $|N_{1/2^+}({\cal M}_0),-\frac{\bm{q}_\bot}{2};\mu^\prime,t{\rangle}$.
The wave function of the final state 
$|N^\prime_{1/2^+}({\cal M^\prime}_0),\frac{\bm{q}_\bot}{2};
\mu^\prime,t{\rangle}$
corresponds to the momentum $+\frac{\bm{q}_\bot}{2}$ and can be obtained from
Eq.{~(\ref{s3st}) by the substitutions 
$\bm{\Lambda}_\bot\to\bm{\Lambda}^\prime_\bot$ and 
${\cal M}_0\to{\cal M}^\prime_0$.

\subsection{Matrix elements of the one-particle current}\label{s23}

We now will have a look at the well-known basic formulas following 
Ref.~\cite{keister88}.
We start with the electromagnetic current of a free quark considered 
as a Dirac particle with charge $e_q$ and anomalous magnetic moment 
$\varkappa_q$ given as
\begin{equation}
j_q^\mu =
e_q\left(\gamma^\mu +\frac{\varkappa_q}{2m}i\sigma^{\mu\nu}q_\nu\right)\, .
\label{jq}
\end{equation}
The $z^+$ component of this current $I_q^+=j_q^0+j_q^3$ plays a decisive role
in the LF approach. As has been shown~\cite{keister88}
in the special Breit frame (\ref{q}), where $q^0=q^3=$0, 
that the matrix element
of any component of the one-particle current (\ref{jq}) can be expressed 
in terms of the $I^+$ matrix element, provided that current conservation 
$j^\mu q_\mu=$0 is obeyed. Hence the nucleon form factors $F_1$ (Dirac) and
$F_2$ (Pauli) can be calculated in terms of matrix elements of the
$I^+$ component of the current
\begin{equation}
I^{(i)+}_q=e^{(i)}_q\left(If_1-i\bm{\sigma}^{(i)}\!\cdot\!
[\bm{z}\times\bm{q}_\bot]\frac{f_2}{2m}\right)
\label{iq}
\end{equation}
and
\begin{equation}
I_{\scriptscriptstyle N}^+=e_{\scriptscriptstyle N}\left(IF_1
-i\bm{\sigma}_{\scriptscriptstyle N}\!\cdot\![\bm{z}\times\bm{q}_\bot]
\frac{F_2}{2m_{\scriptscriptstyle N}}\right)
\label{in}
\end{equation}
written in the special Breit frame (\ref{q}). In both cases
the electric charge (without the factor $e=\sqrt{4\pi\alpha}$)
\begin{equation}
e^{(i)}_q=\frac{1}{6}+\frac{1}{2}\tau^{(i)}_3,\,\,
e_{\scriptscriptstyle N}=\frac{1}{2}+\frac{1}{2}\tau_{{\scriptscriptstyle N}3}
\label{eq}
\end{equation}
is included in the current. Quark form factors $f_1$ and $f_2$ 
could be included in addition, but here we consider the simplest variant 
without quark form factors ($f_1=$1,\,$f_2=\varkappa_q$) assuming that 
the quark is an elementary particle. 

Current matrix elements between LF spin states
\begin{multline}
{_{f}\langle}\frac{1}{2},\bar\mu_3|I^{(3)+}_q|\frac{1}{2},\mu_3\rangle_{f}
=I^{D+}_{\bar\mu_3\mu_3}+I^{P+}_{\bar\mu_3\mu_3}\\
=e^{(3)}_q\left[\delta_{\bar\mu_3,\mu_3}
+\delta_{\bar\mu_3,-\mu_3}(-1)^{1/2-\mu_3}
\frac{q_\bot}{2m}\varkappa_q\right]
\label{iqmu}
\end{multline}
\begin{multline}
{_{f}\langle} N_{1/2^+}^\prime;\bar\mu|I^{\,+}_{\scriptscriptstyle N}
|N_{1/2^+};\mu{\rangle_{f}}\\
=e_{\scriptscriptstyle N}\!\left[\delta_{\bar\mu,\mu}F_1
+\delta_{\bar\mu,-\mu}(-1)^{1/2-\mu}
\frac{q_\bot}{2m_{\scriptscriptstyle N}}F_2
\right]
\label{inmu}
\end{multline}
have a momentum-independent Dirac part 
$I^{D+}_{\bar\mu_3\mu_3}=\delta_{\bar\mu_3\mu_3}$ which only depends 
on the spin indices (for definiteness we take the quark number $i=3$).

The canonical spin matrix elements for the electromagnetic transitions 
$N+\gamma^*\to N^\prime$ ($N^\prime=N,\,N^*$) are determined by the LF matrix
element (\ref{inmu}) using the following decomposition:
\begin{multline}
{_{c}\langle} N^\prime_{1/2^+};\mu^{\prime\prime},t|
{R^{(\scriptscriptstyle N)}_{\scriptscriptstyle M}}^{\,\dagger}
I_{\scriptscriptstyle N}^{\,+}R^{(\scriptscriptstyle N)}_{\scriptscriptstyle M}
|N_{1/2^+};\mu^\prime,t{\rangle_{c}}\\
=\sum_{\bar\mu\mu}{_{c}\langle} N_{1/2^+}^\prime;\mu^{\prime\prime}|
{R^{(\scriptscriptstyle N)}_{\scriptscriptstyle M}}^{\,\dagger}
(\mu^{\prime\prime},\bar\mu)
|N_{1/2^+}^\prime;\bar\mu{\rangle_{f}}\\
\times{_{f}\langle}
 N_{1/2^+}^\prime;\bar\mu,t^\prime|I^{\,+}_{\scriptscriptstyle N}
|N_{1/2^+};\mu,t{\rangle_{f}}\\
\times {_{f}\langle}N_{1/2^+};
\mu|R^{(\scriptscriptstyle N)}_{\scriptscriptstyle M}
(\mu,\mu^\prime)|N_{1/2^+};\mu^\prime{\rangle_{c}}
\label{inc}
\end{multline}
where (see, e.g., the first reference in Ref.~\cite{keister88})
\begin{equation}
R^{(\scriptscriptstyle N)}_{\scriptscriptstyle M}(\mu,\mu^\prime)
=D^{(\frac{1}{2})}_{\mu\mu^\prime}(\theta_M)
\label{rmn}
\end{equation}
\begin{equation}
\cos\frac{\theta_M}{2}=
\frac{1+\sqrt{1+\tau}}{\sqrt{(1+\sqrt{1+\tau}\,)^2+\tau}}\,,
\quad\tau=\frac{Q^2}{4m_{\scriptscriptstyle N}^2}\,.
\label{tm}
\end{equation}
The observed nucleon electric and magnetic form factors are calculated 
with the matrix element~(\ref{inc}). 

Now we define the nucleon current as a sum of single-quark currents
\begin{equation}
I^+_{N(3q)}=\sum_{j=1}^3e^{(j)}_qI^{(j)+}_q
\label{ncur}
\end{equation}
and calculate the nucleon Dirac/Pauli form factors with the quark wave 
functions defined in Eq.~(\ref{s3st}). For $F_1$ and $F_2$ we use the 
following definitions 
(see Refs.~\cite{cardarelli97,Schlumpf95,Schlumpf92} for details)
\begin{eqnarray}
F_1&=&\frac{1}{2}\sum_{\mu^\prime\mu^{\prime\prime}}
\delta_{\mu^\prime,\mu^{\prime\prime}}
\nonumber\\
&{\times}&{_{c}\langle} N^\prime_{1/2^+};\mu^{\prime\prime},t|
{{\cal R}_{\scriptscriptstyle M}}^{\dagger}
I_{{\scriptscriptstyle N}(3q)}^{\,+}{\cal R}_{\scriptscriptstyle M}
|N_{1/2^+};\mu^\prime,t{\rangle_{c}},\nonumber\\
F_2&=&\frac{1}{2}\sum_{\mu^\prime\mu^{\prime\prime}}
\delta_{\mu^\prime,-\mu^{\prime\prime}}
(-1)^{\frac{1}{2}-\mu^\prime}\nonumber\\
&{\times}&{_{c}\langle} N^\prime_{1/2^+};\mu^{\prime\prime},t|
{{\cal R}_{\scriptscriptstyle M}}^{\dagger}
I_{{\scriptscriptstyle N}(3q)}^{\,+}{\cal R}_{\scriptscriptstyle M}
|N_{1/2^+};\mu^\prime,t{\rangle_{c}}.
\label{f3q}
\end{eqnarray}
Here ${\cal R}_{\scriptscriptstyle M}$ is the three-quark Melosh rotation 
defined in Eqs.~(\ref{mel})-(\ref{3mel}). Further on the standard relation 
between the Sachs and Dirac/Pauli form factors
\begin{equation}
G_E=F_1-\tau F_2,\quad G_M=F_1+F_2
\label{sacs}
\end{equation}
can be used.

\subsection{Form factors in terms of the single-quark current}\label{s24}

To calculate the nucleon form factors $F_1$,$\,F_2$ 
when starting from the LF quark
current (\ref{iqmu}) we define the nucleon matrix 
element~(\ref{inc}) in terms of the quark wave function (\ref{s3st}) 
deduced in Sec.~\ref{s22}. The definition of the quark matrix element implies 
integration over the wave functions involving the six-dimensional momentum
space
\begin{multline}
{_{c}\langle} N^\prime_{1/2^+},\frac{\bm{q}_\bot}{2};\mu^{\prime\prime},t|
{{\cal R}_{\scriptscriptstyle M}}^{\,\dagger}
I_{{\scriptscriptstyle N}(3q)}^{\,+}{\cal R}_{\scriptscriptstyle M}
|N_{1/2^+},-\frac{\bm{q}_\bot}{2};\mu^\prime,t{\rangle_{c}}\\
=\frac{{\cal N}_p}{(2\pi)^6}\int_0^1d\xi\int_0^1d\eta\int d^2\bm{\Lambda}_\bot
\int d^2\bm{\lambda}_\bot
J(\xi,\eta,\bm{\Lambda}_\bot,\bm{\lambda}_\bot)\\
\times 3\langle N^\prime_{1/2^+}({\cal M}^\prime_0);\mu^{\prime\prime},t|
I^{(3)+}_q|N_{1/2^+}({\cal M}_0);\mu^\prime,t{\rangle}\,.
\label{int}
\end{multline}
The integrand in Eq.~(\ref{int}) includes the combinatorial factor 3 
(the number of quarks in the system) and the Jacobian $J$ which corresponds to 
the transition from ordinary quark momenta $\bm{k}_1,\bm{k}_2,\bm{k}_s$ to the 
relative LF variables~(\ref{lam}) and~(\ref{xi}) 
\begin{equation}
J(\xi,\eta,\bm{\Lambda}_\bot,\bm{\lambda}_\bot)=\frac{P^+}{{P^\prime}^+}
\frac{\sqrt{\prod_{i=1}^3\omega_i\prod_{j=1}^3\omega^\prime_j}}
{\xi(1-\xi)\eta(1-\eta)\sqrt{{\cal M}_0{\cal M}^\prime_0}}
\label{jac}
\end{equation}
Here a standard definition for the $i$th quark energy $\omega_i$
is used 
\begin{eqnarray}
\omega_i&=&\frac{1}{2}\left({\cal M}_0 x_i+
\frac{m^2+\bm{k}^2_{\bot i}}{{\cal M}_0 x_i}\right),\,\,
\bm{k}_{\bot 1}=\bm{\lambda}_\bot+\xi\bm{\Lambda}_\bot,\nonumber\\
\bm{k}_{\bot 2}&=&-\bm{\lambda}_\bot+(1-\xi)\bm{\Lambda}_\bot,\,\,
\bm{k}_{\bot 3}=-\bm{\Lambda}_\bot
\label{omg}
\end{eqnarray}
while the notation $\omega^\prime_j$ is reserved for the energy of $j$th 
quark in the final state [then the substitutions
$\bm{k}_{\bot i}\to$ $\bm{k}^\prime_{\bot j}$, $\,\bm{\Lambda}_\bot\to$
$\bm{\Lambda}^\prime_\bot$ and ${\cal M}_0\to$ ${\cal M}^\prime_0$
should be made in Eq.~(\ref{omg})]. 
The matrix element (\ref{int}) is normalized with the factor ${\cal N}_p$ 
which provides the correct proton charge of unity, i.e $F_{1p}(0)=$1. 

The integrand in Eq.~(\ref{int}) consists of two components which originate 
from the Dirac ($D$) and Pauli ($P$) terms of the quark current of 
Eq.~(\ref{iqmu}). 
For the Dirac component of the integrand we use the representation
\begin{multline}
{\cal J}_{\mu^{\prime\prime}\mu^{\prime}}^D({\cal M}^\prime_0,{\cal M}_0;t)\\
=3\langle N^\prime_{1/2^+}({\cal M}^\prime_0);\mu^{\prime\prime},t|
\sum_{\mu_3\bar\mu_3}|\frac{1}{2},\bar\mu_3
\rangle\langle\frac{1}{2},\bar\mu_3|\\
\times I^{D+}_{\bar\mu_3\mu_3}|\frac{1}{2},\mu_3\rangle\langle\frac{1}{2},\mu_3
|N_{1/2^+}({\cal M}_0);\mu^\prime,t{\rangle}
\label{id}
\end{multline}
and the same formula with the substitution
$I^{D+}_{\bar\mu_3\mu_3}\to$ $I^{P+}_{\bar\mu_3\mu_3}=(\varkappa_q q_\bot/2m)
\delta_{\bar\mu_3,-\mu_3}(-1)^{1/2-\mu_3}$ is used for the Pauli component
${\cal J}_{\mu^{\prime\prime}\mu^{\prime}}^P({\cal M}^\prime_0,{\cal M}_0;t)$.

It is rather straightforward to derive explicit expressions for 
${\cal J}_{\mu^{\prime\prime}\mu^{\prime}}^D$ and 
${\cal J}_{\mu^{\prime\prime}\mu^{\prime}}^P$ in terms of the coefficients 
$C^{SS^\prime}_{S_{12},S_{12}^\prime}$ when the wave functions
$|N_{1/2^+}({\cal M}_0);\mu^\prime,t{\rangle}$ and 
$|N^\prime_{1/2^+}({\cal M}^\prime_0);\mu^{\prime\prime},t{\rangle}$ 
are substituted in the form given in Eq.~(\ref{s3st}) with 
the basis vectors (\ref{s3s1}) and (\ref{s3s2}) defined in Sec.~\ref{s21}.
The result is
\begin{multline}
{\cal J}_{\mu^{\prime\prime}\mu^{\prime}}^D({\cal M}^\prime_0,{\cal M}_0;t)
=3\Phi_{S^\prime}({\cal M}^\prime_0)\Phi_S({\cal M}_0)
\frac{1}{2}\sum_{\mu\bar\mu}
\delta_{\mu\bar\mu}\\
\times\!\left\{\!\left[
{{C}^\prime}^*_{01}(\bar\mu,\mu^{\prime\prime})
{C}_{01}(\mu,\mu^\prime)
+{C^\prime}^*_{11}(\bar\mu,\mu^{\prime\prime})C_{11}(\mu,\mu^\prime)
\right]\!\!e_1(t)\right.\\
\left.+\!\left[
{C^\prime}^*_{00}(\bar\mu,\mu^{\prime\prime})C_{00}(\mu,\mu^\prime)
+{{C}^\prime}^*_{10}(\bar\mu,\mu^{\prime\prime})
{C}_{10}(\mu,\mu^\prime)
\right]\!\!e_2(t)\!\right\},
\label{jd}
\end{multline}
\begin{multline}
{\cal J}_{\mu^{\prime\prime}\mu^{\prime}}^P({\cal M}^\prime_0,{\cal M}_0;t)\\
=3\Phi_{S^\prime}({\cal M}^\prime_0)\Phi_S({\cal M}_0)\frac{1}{2}
\frac{\varkappa_q q_\bot}{2m}\sum_{\mu\bar\mu}A_{\bar\mu,\mu}\\
\times\!\left\{\!\left[
{{C}^\prime}^*_{01}(\bar\mu,\mu^{\prime\prime})
{C}_{01}(\mu,\mu^\prime)
+{C^\prime}^*_{11}(\bar\mu,\mu^{\prime\prime})C_{11}(\mu,\mu^\prime)
\right]\!\!e_1(t)\right.\\
\left.+\!\left[
{C^\prime}^*_{00}(\bar\mu,\mu^{\prime\prime})C_{00}(\mu,\mu^\prime)
+{{C}^\prime}^*_{10}(\bar\mu,\mu^{\prime\prime})
{C}_{10}(\mu,\mu^\prime)\right]\!\!e_2(t)\!\right\},
\label{jp}
\end{multline}
where the matrix $A_{\bar\mu,\mu}$ is given in the Appendix and each term  
${{C}^\prime}^*_{S_{12}S^\prime_{12}}(\bar\mu,\mu^{\prime\prime})
{C}_{S_{12}S^\prime_{12}}(\mu,\mu^\prime)$ in the rhs of Eqs.~(\ref{jd}) and 
(\ref{jp}) is a symbolical expression that implies a sum of two terms for
the front spins $S=\frac{1}{2}$ and $\frac{3}{2}$ as it follows from 
Eqs.~(\ref{s3s1}) and (\ref{s3s2}), e.g., 
\begin{multline}
{{C}^\prime}^*_{11}(\bar\mu,\mu^{\prime\prime})
{C}_{11}(\mu,\mu^\prime)\doteq\\
{C^{\frac{1}{2}\frac{1}{2}}_{11}}^*(\bm{\lambda}_\bot,
\bm{\Lambda}^\prime_\bot;
\bar\mu,\mu^{\prime\prime})
C^{\frac{1}{2}\frac{1}{2}}_{11}(\bm{\lambda}_\bot,\bm{\Lambda}_\bot;
\mu,\mu^\prime)\\
+{C^{\frac{3}{2}\frac{1}{2}}}^*_{11}(\bm{\lambda}_\bot,
\bm{\Lambda}^\prime_\bot;
\bar\mu,\mu^{\prime\prime})
C^{\frac{3}{2}\frac{1}{2}}_{11}(\bm{\lambda}_\bot,
\bm{\Lambda}_\bot;\mu,\mu^\prime)
\label{jds}
\end{multline}

In Eqs.~(\ref{jd}) and (\ref{jp}) the isospin factors are
\begin{multline}
e_1(t)=\\
\langle[21]_{\scriptscriptstyle T}y^{(1)}_{\scriptscriptstyle T}\!\!,t|
e_q^{(3)}|[21]_{\scriptscriptstyle T}
y^{(1)}_{\scriptscriptstyle T}\!\!,t{\rangle}=
\begin{cases}
0, &t=+\frac{1}{2},\\
\frac{1}{3},&t=-\frac{1}{2}
\end{cases}
\label{et1}
\end{multline}
\begin{multline}
e_2(t)=\\
\langle[21]_{\scriptscriptstyle T}y^{(2)}_{\scriptscriptstyle T}\!\!,t|
e_q^{(3)}|[21]_{\scriptscriptstyle T}
y^{(2)}_{\scriptscriptstyle T}\!\!,t{\rangle}=
\begin{cases}
\frac{2}{3}, &t=+\frac{1}{2},\\
-\frac{1}{3},&t=-\frac{1}{2}
\end{cases}
\label{et2}
\end{multline}

In the following $F^D_{1t}$ and $F^P_{1t}$ denote the contributions 
of the Dirac and Pauli 
quark currents $I^{D+}_{\bar\mu_3\mu_3}$ and $I^{P+}_{\bar\mu_3\mu_3}$ to the 
nucleon $F_1$ form factor: $F_{1t}=F^D_{1t}+F^P_{1t}$, 
while the same notations 
are used for the nucleon $F_2$ form factor: $F_{2t}=F^D_{2t}+F^P_{2t}$.
These contributions to $F_{1t}(F_{2t})$ are represented by the following 
six-dimensional integrals of the functions defined in Eqs.~(\ref{jd}) 
and (\ref{jp}) with 
\begin{equation}
F^D_{1t}=\int\! d{\cal V}_{\scriptscriptstyle LF}\frac{1}{2}
\sum_{\mu^\prime\mu^{\prime\prime}}
\delta_{\mu^\prime,\mu^{\prime\prime}}
{\cal J}_{\mu^{\prime\prime}\mu^{\prime}}^D({\cal M}^\prime_0,{\cal M}_0;t)
\label{fd1}
\end{equation}
\begin{multline}
F^D_{2t}=\int\! d{\cal V}_{\scriptscriptstyle LF}\frac{1}{2}
\sum_{\mu^\prime\mu^{\prime\prime}}
\delta_{\mu^\prime,-\mu^{\prime\prime}}(-1)^{1/2-\mu^\prime}\\
\times\frac{2m_N}{q_\bot}
{\cal J}_{\mu^{\prime\prime}\mu^{\prime}}^D({\cal M}^\prime_0,{\cal M}_0;t)
\label{fd2}
\end{multline}
\begin{multline}
F^P_{1t}=\int\! d{\cal V}_{\scriptscriptstyle LF}\frac{1}{2}
\sum_{\mu^\prime\mu^{\prime\prime}}
\delta_{\mu^\prime,\mu^{\prime\prime}}
{\cal J}_{\mu^{\prime\prime}\mu^{\prime}}^P({\cal M}^\prime_0,{\cal M}_0;t)
\label{fp1}
\end{multline}
\begin{multline}
F^P_{2t}=\int\! d{\cal V}_{\scriptscriptstyle LF}\frac{1}{2}
\sum_{\mu^\prime\mu^{\prime\prime}}
\delta_{\mu^\prime,-\mu^{\prime\prime}}(-1)^{1/2-\mu^\prime}\\
\times\frac{2m_N}{q_\bot}
{\cal J}_{\mu^{\prime\prime}\mu^{\prime}}^P({\cal M}^\prime_0,{\cal M}_0;t)\,.
\label{fp2}
\end{multline}
Here we denote the integration volume in compact form as
\begin{equation*}
d{\cal V}_{\scriptscriptstyle LF}=J(\xi,\eta,\bm{\Lambda}_\bot,
\bm{\lambda}_\bot)\,
d\xi d\eta d^2\Lambda_\bot d^2\lambda_\bot \,.
\label{vol}
\end{equation*} 

\section{Description of data on form factors and helicity amplitudes}
\label{s3}

\subsection{Nucleon form factors}\label{s31}

Previous results enable us to determine the nucleon form factors in a wide 
$Q^2$ range from 0 to 35 GeV$^2$. The nucleon form factors $F_1/F_2$ are 
defined as the sums of matrix elements (\ref{fd1})--(\ref{fp2}) of the 
Dirac/Pauli quark currents 
\begin{equation}
F_{1t}=F^D_{1t}+F^P_{1t},\quad F_{2t}=F^D_{2t}+F^P_{2t}
\label{f1f2}
\end{equation}
where $t=+$1/2 (proton), $t=-$1/2 (neutron). The Sachs form factors $G_E/G_M$
are defined by Eq.~(\ref{sacs}). 

In the LF approaches~\cite{capstick95,cardarelli97,aznauryan12,Schlumpf95}
the ``radial'' part of the $S$-wave quark core of the nucleon is usually 
described by the Gaussian~$\Phi_{0S}={\cal N}_{0S}\, 
\exp(-{\cal M}^2_0/2\beta^2)$. 
However, for large $Q^2$ the elastic or 
inelastic form factors are suppressed by the Gaussian, 
hence the pQCD prediction
for their asymptotic behavior with $G_E\sim 1/Q^4$, $G_M\sim 1/Q^6$ cannot
be provided. 

A superposition of several harmonic oscillator wave functions (up to 20) was 
used in more advanced approaches~\cite{cardarelli97,capstick07} to obtain a 
realistic description of form factors at low and moderate values of $Q^2$. 
Nevertheless, the problem of the asymptotic power behavior of nucleon form 
factors can only be solved in such models where many free parameters are 
available to be fitted to the data up to high values of $Q^2$. 

Recently, in Ref.~\cite{aznauryan12}, a running quark mass [in the pole 
form $m(Q^2)=m(0)/(1+Q^2/\Lambda^2)$] was used in the LF model 
with a Gaussian shaped wave function. This model is consistent (at least 
qualitatively) with the QCD prediction for the $Q^2$ behavior of the quark 
mass. The problem is essentially as follows: in quark models the factor 
$m^{-1}(Q^2)$ appears in the expression for the nucleon form factor $F_2$. 
Therefore, the falloff of the Gaussian in $F_2$ at large $Q^2$ can be 
compensated by an increase in $m^{-1}(Q^2)$. As a result a realistic 
description of the $Q^2$ behavior of both the nucleon form factors and the 
Roper resonance production helicity amplitudes was obtained in 
Ref.~\cite{aznauryan12}.

Here we suggest an alternative method for solving the problem of the
``non-Gaussian behavior'' of form factors at large $Q^2$. In our opinion a
pole form of the nucleon/Roper wave function
\begin{equation}
\Phi_{0S}={\cal N}_{0S}\frac{1}{(1+{\cal M}_0^2/\beta^2)^\gamma}
\label{pol}
\end{equation}
is also workable and correlates well with QCD predictions for the high $Q^2$ 
behavior of elastic and inelastic form factors. 
Two decades ago such a pole form for the nucleon wave function 
was considered in Ref.~\cite{Schlumpf92}. A realistic description of 
nucleon form factors and of magnetic moments was obtained for a value of 
$\gamma=3.5$. At this time the theoretical description of data on $G_E,\,G_M$ 
was rather reasonable; the case of the Roper 
resonance was not discussed because of the absence of good data.
Now a large set of new high-quality data  both on the nucleon form 
factors~\cite{punjabi05,puckett12,ron11,kubon02,xu00,xu03,milbrath98,%
jones00,crowford07,jones06,maclachlan06,puckett10} and on the 
electroproduction of the Roper 
resonance~\cite{aznauryan09,mokeev09,sarantsev08} are available. 
Hence a precise analysis in terms of a common approach both to elastic and 
inelastic form factors is possible now..

Here we show that the given LF quark model allows for a good description 
of all the new data on nucleon form factors in a large interval of $Q^2$ from
0 up to 35 GeV$^2$. The model has only five free parameters 
[see Eqs.~(\ref{jq}) and~(\ref{pol})], 
$\gamma$, $\beta$, $m$, and $\varkappa_q$, 
which are fitted to the data. For the values 
\begin{multline}
\gamma=\mbox{3.51},\,\,
\beta=\mbox{0.579 GeV},\,\, m=\mbox{0.251 GeV},\\ 
\varkappa_u=-\mbox{0.017},\,\,\varkappa_d=\mbox{0.0295}
\label{param}
\end{multline}
an optimal description of the elastic nucleon data is obtained 
(see Figs.~\ref{f1}--\ref{f5n}). We present a comparison with known data 
and the soft-wall AdS/QCD approach~\cite{Gutsche:2012bp}, where elastic 
nucleon form factors and nucleon-Roper transition form factors have been 
analyzed in detail. 
The model generates a $Q^2$ behavior for
$Q^4F_{1p}(Q^2)$, $Q^4F_{1n}(Q^2)$ and $Q^2F_{2p}(Q^2)/F_{1p}(Q^2)$ 
which should tend to a constant at high $Q^2$ 
(see Figs. \ref{f1}, \ref{f1n} and \ref{f3}). 
At low and moderate values of $Q^2$ the model is
compatible not only with the magnetic moments of the nucleons,
$\mu_p=F_{1p}(0)+F_{2p}(0)=$2.79 and $\mu_n=F_{1n}(0)+F_{2n}(0)=-$1.91 but also
with the known negative slope for the ratio $G^p_E(Q^2)/G^p_M(Q^2)$ 
(Fig. \ref{f4}). The absolute theory values for $G^p_M(Q^2)$, $G^n_M(Q^2)$ and
$G^p_E(Q^2)$ do also correlate well with the data as evident from 
Figs.~\ref{f2}, \ref{f2n} and \ref{f5}, respectively, where the 
dipole form factor $G_D=(1+Q^2/0.71)^{-2}$ is used as a common denominator. 

Only in the case of the neutron charge form factor $G^n_E$ (Fig.~\ref{f5n})  
this model is not entirely adequate to describe data at low and 
intermediate values of $Q^2$. But in this $Q^2$ region the pion cloud 
contribution to $G^n_E$, neglected in the present work, can be considerable. 
This contribution can also be 
important for inelastic nucleon form factors at $Q^2\lesssim$1 GeV$^2$, as was 
recently noted in Ref.~\cite{aznauryan12}. In our recent 
work~\cite{obukhovsky11} we also pointed out the role of the pion cloud in the 
electroproduction of the Roper resonance. 

\subsection{Helicity amplitudes in the electroexcitation of the 
Roper resonance}
\label{s32}

As was suggested in Refs.~\cite{obukhovsky11,aznauryan12} the 
$\sigma$ meson  along with the pion cloud can contribute significantly to 
the process $p+\gamma^*\to R$. In Ref.~\cite{obukhovsky11} we considered
the contribution of the $\sigma $ to the electroproduction
of the Roper resonance by assuming a composite structure.
In this case the Roper resonance is set up as a superposition 
of the radially excited three-quark configuration $3q^\ast$ and 
the hadron molecule component $N+\sigma$ as 
\begin{equation} 
|R\ra = \cos\theta |3q^\ast\ra + \sin\theta |N+\sigma\ra\,.
\label{theta}
\end{equation}
A mixing angle $\theta$ is introduced:
$\cos^2\theta $ and $\sin^2\theta$ represent the 
probabilities to find a $3q^\ast$ and hadronic configuration, respectively.
The parameter $\theta$ was adjusted to optimize the description of the
electroproduction data resulting in the optimal value of
$\cos\theta=$0.8~\cite{obukhovsky11}.

In Ref.~\cite{obukhovsky11} the contribution of a hadronic $N + \sigma$ 
component to the process $p+\gamma^*\to R$ was calculated in the framework of 
a relativistic approach; the related technique was proposed and extensively 
used in Ref.~\cite{RQM}.  
The interaction vertices of these diagrams are derived using 
nonlocal relativistic Lagrangians with $NN\sigma$ and $NR\sigma$ 
couplings, which are manifestly gauge invariant. 

But at the same time a somewhat inconsistent (nonrelativistic) 
technique was used for the quarks. The contribution of the nucleon quark 
core to the transition $3q+\gamma^*\to 3q^\ast$ was calculated in terms of 
a nonrelativistic quark shell model with Gaussian wave functions. 
Here we remedy this original defect and recalculate the quark amplitudes 
with the quark LF wave functions of the nucleon, Eq.~(\ref{pol}), 
and the Roper resonance, Eq.~(\ref{phi02s}).

The transverse and longitudinal helicity amplitudes $A_{1/2}$ and $S_{1/2}$ 
for the electroproduction
of the Roper resonance are defined (see, e.g. Ref.~\cite{tiator09}) by
matrix elements of the hadronic current $J^\mu$. These matrix elements
\begin{multline}
\langle R(p^\prime,\lambda^\prime)|J^\mu(0)|N(p,\lambda)\rangle\\
=\bar u_R(p^\prime,\lambda^\prime)\left\{F_1^{RN}(Q^2)
\left(\gamma^\mu-\not\! q\frac{q^\mu}{q^2}\right)\right.\\
\left.+F_2^{RN}(Q^2)\frac{i\sigma^{\mu\nu}q_\nu}{m_R+m_N}u_N(p,\lambda)\right\}
\label{jrn}
\end{multline}
are taken between the helicity states of the initial
nucleon $|N(p,\lambda)\rangle$ and the final Roper resonance 
$|R(p^\prime,\lambda^\prime)\rangle$
(we use notations and definitions given in Ref.~\cite{tiator09}, where
formulas are written in the rest frame of the Roper resonance) with
\begin{equation}
A_{1/2}=-\sqrt{\frac{\pi\alpha}{2k_{\scriptscriptstyle R}m_Rm_N}}
\langle R(p^\prime,\frac{1}{2})|\bm{J}\!\cdot\!\bm{\epsilon}_{\lambda=+1}
|N(p,-\frac{1}{2})\rangle
\label{a12}
\end{equation}
\begin{equation}
S_{1/2}=-\sqrt{\frac{\pi\alpha}{2k_{\scriptscriptstyle R}m_Rm_N}}
\langle R(p^\prime,\frac{1}{2})|J^0|N(p,\frac{1}{2})\rangle
\label{s12}
\end{equation}
where $k_{\scriptscriptstyle R}=(m_R^2-m_N^2)/(2m_R)$ and the transverse 
polarization vector is 
$\{\b{\epsilon}_{\lambda=+1}\}=-\sqrt{\frac{1}{2}}\{1,i,0\}$.

It follows from Eqs.~(\ref{jrn})--(\ref{s12}) that the amplitudes 
$A_{1/2}$ and $S_{1/2}$ can be expressed in terms of the invariant 
form factors $F_1^{RN}(Q^2)$ and $F_2^{RN}(Q^2)$ for which we already 
have explicit expressions~(\ref{fd1})-(\ref{f1f2}) in the LF formalism. 
We have the relations
\begin{equation}
A_{1/2}=\sqrt{\frac{\pi\alpha}{k_{\scriptscriptstyle R}m_Rm_N}}
Q_-(F_1^{RN}+F_2^{RN}),
\label{afrn}
\end{equation}
\begin{multline}
S_{1/2}=\sqrt{\frac{\pi\alpha}{2k_{\scriptscriptstyle R}m_Rm_N}}
\frac{Q_+Q_-}{Q^2}\frac{m_R+m_N}{2m_R}\\
\times Q_-\left(F_1^{RN}-\frac{Q^2}{(m_R+m_N)^2}F_2^{RN}\right),
\label{sfrn}
\end{multline}
where $Q_\pm=\sqrt{(m_R\pm m_N)^2+Q^2}$.
$F_1^{RN}(Q^2)$ and $F_2^{RN}(Q^2)$ are given by 
Eqs.~(\ref{fd1})--(\ref{f1f2}) 
with the wave functions $\Phi_S=\Phi_{0S}$ (for the initial nucleon) and 
$\Phi_{S^\prime}=\Phi_{2S}$ (for the final Roper resonance).

The calculated helicity amplitudes $A_{1/2}$ and $S_{1/2}$ are shown in 
Fig.~\ref{fr12} for unchanged values of the model parameters, 
Eq.~(\ref{param}),
previously fitted to the nucleon data. The only additional free parameter 
is the mixing angle $\theta $ of Eq.~(\ref{theta}). 
We vary $\cos\theta$ from 0.7 to 1 where
results for the specific values $\cos\theta=1$ 
and 0.7 are shown in Fig.~\ref{fr12}.
When the Roper wave function corresponds to a 
$3q^*$ state ($\cos\theta=$1) predictions for both helicity 
amplitudes $A_{1/2}$ and $S_{1/2}$ (dashed lines)
are much too large in comparison
with the data~\cite{aznauryan09,mokeev09}. A value of about
$\cos\theta=$0.7 (solid lines) is preferred in the present model
where data for $Q^2\lesssim$2~GeV$^2$ of both helicity amplitudes can be
roughly reproduced. For $Q^2$ in the range from 2 to 4 GeV$^2$ the behavior
of  the $A_{1/2}$ data cannot be explained sufficiently. 
Note that in our previous nonrelativistic model~\cite{obukhovsky11} 
(dashed dotted lines in Fig. \ref{fr12}) a better description of the data 
was achieved with an optimal value of $\cos\theta=$0.8.

For comparison we also show
in Fig.~\ref{fr12} the results of Ref.~\cite{aznauryan12} for the LF model with
running quark masses (the double-dotted dashed line). 
The weight of the three-quark core in the Roper wave function
in that model of $\cos\theta=\,$0.73 is very close to the value 
$\cos\theta=\,$0.7
deduced in our model. Both predictions have a similar behavior for
$A_{1/2}$ at low and intermediate values of $Q^2\lesssim\,$4--5 GeV$^2$ 
but at higher $Q^2$ the predictions increase relative to present data. 
On the basis of the CLAS data at intermediate 
$Q^2\lesssim\,$4--5 GeV$^2$
both models point to the same (about 50\%) probability for the $3q$ component 
in the wave function of the Roper resonance. We can conclude that the quark LF 
model might represent a viable approach to the description of the data at high
$Q^2$ if the $3q^*$ component in the Roper resonance is subleading, maximally
with a probability of not even 50 \%. The remaining structure is dominated by 
more complicated soft components: the meson cloud or 
$q\bar q$, $qq\bar q\bar q$, etc. states. The contribution of these 
configurations to form factors becomes negligible with growing $Q^2$. Soft 
components can also modify the $Q^2$ behavior of the helicity amplitudes at 
low and moderate values of $Q^2$, while in the high $Q^2$ region the $3q^*$ 
core defines the power asymptotics of the amplitudes predicted by pQCD.
 
\section{Summary}\label{s4}

We presented a version of the light-front approach to elastic and inelastic
nucleon form factors in which the LF three-quark configurations (obtained 
with the Melosh rotation of the canonical spin states 
$|s_1s_2(S_{12})s_3:S,\mu{\rangle}_c$)
satisfy the Pauli exclusion principle on the LF.
Such an approach is equivalent to a variant where configuration 
mixing on the LF is neglected, 
but it appears to be a good approximation at least for nucleons. Both
$N+\gamma^*\to N$ and $N+\gamma^*\to N^*$ transition amplitudes are 
described in a common framework based on a 
relativistic quark model satisfying the Pauli exclusion principle on the LF. 

A pole form for the ``radial'' part of the nucleon wave function allows
for a good description of the nucleon form factors in a large $Q^2$ 
region from  $Q^2=$0 (proton and neutron magnetic moments) to high values
of about 30 GeV$^2$ (with the power behavior $\sim \, Q^{-4}$, 
$\sim \, Q^{-6}$ for $G_E$, $G_M$). 
At the same time the calculated helicity amplitudes $A_{1/2}(Q^2)$ and
$S_{1/2}(Q^2)$ of electroproduction of the Roper resonance on the proton 
occur to be  too large in comparison with
recent data at low and moderate values of $Q^2$. A closer description of 
both $A_{1/2}(Q^2)$ and $S_{1/2}(Q^2)$ amplitudes
can only be obtained if the quark core configuration in the Roper resonance
is suppressed.
The remaining part of  the full $N^*$ state could be a soft component which is
described in terms of a meson cloud or equivalently in terms of a soft cloud 
of $q\bar q$, $(qq)(\bar q\bar q)$,\dots etc. pairs.

In a first approximation we assume that this soft component can be 
described by a hadronic molecular $N+\sigma$ state, the inner structure of
which we have studied in our recent work~\cite{obukhovsky11}. Our evaluation 
shows that the contribution of such a component to the inelastic $N\to R$ form 
factors $F^{RN}_1(Q^2)$ and  $F^{RN}_2(Q^2)$ becomes negligible with growing
$Q^2$. The asymptotic $Q^2$ behavior of $F^{RN}_1$ and  $F^{RN}_2$ is 
given by the quark core component $3q^*$ resulting in a true power law 
and with a reduced absolute value of the amplitude (about 50\% compared 
to the full $3q\to 3q^*$ amplitude). At low and moderate values of $Q^2$
the hadronic molecule component improves the description of the form factors.
At this level the considered model is too simple to provide strong evidence 
for a large hadronic $N+\sigma$ component which 
effectively represents the soft part of the full wave function. 

\vspace*{.5cm}

\begin{acknowledgments}  

This work was supported by the DFG under Contracts No. FA67/39-1 
and No. LY 114/2-1. The work is done partially under  
Project No. 2.3684.2011 of Tomsk State University. 
V. E. L. would like to thank Institute of Nuclear Physics, Moscow
State University, Russia and Tomsk Polytechnic University, Russia 
for warm hospitality. 

\end{acknowledgments}



\appendix

\section{Coefficients $C^{SS^\prime}_{S_{12},S^\prime_{12}}$}
The coefficients $C^{SS^\prime}_{S_{12},S^\prime_{12}}$ defined in 
Sec.\ref{s21} 
with Eq.~(\ref{fc}) have the following explicit expression
\begin{multline}
C^{\frac{1}{2}\frac{1}{2}}_{00}(\bm{\lambda}_\bot,\bm{\Lambda}_\bot;
\mu,\mu^\prime)
=\frac{N_0}{D}\biggl(\delta_{\mu,\mu^\prime}(m+(1\!-\!\eta){\cal M}_0)\\
\left.-\delta_{\mu,-\mu^\prime}(\Lambda_1(-1)^{\frac{1}{2}
-\mu^\prime}+i\Lambda_2)
\right)
\label{c00}
\end{multline}
\begin{multline}
{C^{\frac{1}{2}\frac{1}{2}}_{00}}^*(\bm{\lambda}_\bot,\bm{\Lambda}^\prime_\bot;
\bar\mu,\mu^{\prime\prime})
=\frac{N_0}{D}\biggl(\delta_{\bar\mu,\mu^{\prime\prime}}(m+(1\!-\!\eta)
{\cal M}^\prime_0)\\
\left.-\delta_{\bar\mu,-\mu^{\prime\prime}}
(\Lambda^\prime_1(-1)^{\frac{1}{2}-\mu^{\prime\prime}}
-i\Lambda^\prime_2)\right)
\label{c00p}
\end{multline}
\begin{multline}
C^{\frac{1}{2}\frac{1}{2}}_{10}(\bm{\lambda}_\bot,\bm{\Lambda}_\bot;
\mu,\mu^\prime)=
-\frac{1}{D}\frac{1}{\sqrt{3}}
\biggl\{(m+(1\!-\!\eta){\cal M}_0)\\
\times\delta_{\mu,-\mu^\prime}\biggl[
(2m+\eta{\cal M}_0)(\lambda_1(-1)^{\frac{1}{2}-\mu^\prime}+i\lambda_2)\\
-(1-2\xi)m(\Lambda_1(-1)^{\frac{1}{2}-\mu^\prime}+i\Lambda_2)\biggl]+\\
(2m+\eta{\cal M}_0)\delta_{\mu,\mu^\prime}\!
\biggl[\bm{\lambda}_\bot\!\cdot\!\bm{\Lambda}_\bot
+i[\bm{\lambda}_\bot\!\times\!
\bm{\Lambda}_\bot]_3(-1)^{\frac{1}{2}-\mu^\prime}\!
\biggl]\\
+\delta_{\mu,\mu^\prime}(1-2\xi)m{\bm{\Lambda}_\bot}^{\!\!2}
\biggl\}
\label{c012}
\end{multline}
\begin{multline}
C^{\frac{3}{2}\frac{1}{2}}_{10}(\bm{\lambda}_\bot,\bm{\Lambda}_\bot;
\mu,\mu^\prime)=\\
\frac{1}{D}\frac{1}{\sqrt{2}}\biggl\{
\frac{1}{\sqrt{3}}\delta_{\mu,-\mu^\prime}(m+(1\!-\!\eta){\cal M}_0)\\
\times\biggl[(2m+\eta{\cal M}_0)
(\lambda_1+i\lambda_2(-1)^{\frac{1}{2}-\mu^\prime})\\
-(1-2\xi)m(\Lambda_1+i\Lambda_2(-1)^{\frac{1}{2}-\mu^\prime})\biggl]+\\
(m+(1\!-\!\eta){\cal M}_0)\biggl[(2m+\eta{\cal M}_0)
(\lambda_1\delta_++i\lambda_2\delta_-\!)\\
-(1-2\xi)m(\Lambda_1\delta_++i\Lambda_2\delta_-\!)\biggl]-(2m+\eta{\cal M}_0)\\
\times\biggl[\frac{1}{\sqrt{3}}\delta_{\mu,\mu^\prime}
\biggl(\bm{\lambda}_\bot\!\cdot\!\bm{\Lambda}_\bot(-1)^{\frac{1}{2}-\mu^\prime}
+i[\bm{\lambda}_\bot\!\times\!\bm{\Lambda}_\bot]_3\biggl)\\
+(\lambda_1\Lambda_1-\lambda_2\Lambda_2)\Delta_-
+i(\lambda_1\Lambda_2+\lambda_2\Lambda_1)\Delta_+\biggl]\\
+(1-2\xi)m\biggl[\frac{1}{\sqrt{3}}
\delta_{\mu,\mu^\prime}(-1)^{\frac{1}{2}-\mu^\prime}
{\bm{\Lambda}_\bot}^{\!\!2}\\
+(\Lambda^2_1-\Lambda^2_2)\Delta_-+2i\Lambda_1\Lambda_2\Delta_+\biggl]
\biggl\}
\label{c032}
\end{multline}
\begin{multline}
C^{\frac{1}{2}\frac{1}{2}}_{11}(\bm{\lambda}_\bot,\bm{\Lambda}_\bot;
\mu,\mu^\prime)=\\
\frac{1}{D}\biggl\{\delta_{\mu,\mu^\prime}\biggl[(m+(1\!-\!\eta){\cal M}_0)
(u-\frac{1}{3}\bm{W}_\bot\!\cdot\!\bm{V}_\bot)\\
-\frac{2}{3}\bm{\Gamma}_\bot\!\cdot\!\bm{\Lambda}_\bot
-\frac{i}{3}(m+(1\!-\!\eta){\cal M}_0)\Gamma_2\biggl]\\
+\delta_{\mu,\mu^\prime}(-1)^{\frac{1}{2}-\mu^\prime}
\biggl[\frac{1}{3}(m+(1\!-\!\eta){\cal M}_0)\Gamma_1
+\frac{2i}{3}\Gamma_2\Lambda_1\biggl]\\
+\delta_{\mu,-\mu^\prime}\biggl[\frac{2}{3}
(m+(1\!-\!\eta){\cal M}_0)\Gamma_2+\frac{2i}{3}\Lambda_1
[\bm{W}_\bot\!\times\!\bm{V}_\bot]_3\\
+i\bigl(\frac{1}{3}\Lambda_2(u-\bm{W}_\bot\!\cdot\!\bm{V}_\bot)
-\frac{2}{3}\Lambda_2(W_1V_1-W_2V_2)\bigl)\biggl]\\
+\delta_{\mu,-\mu^\prime}(-1)^{\frac{1}{2}-\mu^\prime}
\biggl[\frac{2}{3}(m+(1\!-\!\eta){\cal M}_0)\Gamma_1\\
+\frac{1}{3}\Lambda_1(u-\bm{W}_\bot\!\cdot\!\bm{V}_\bot)
-\frac{2}{3}\Lambda_1(W_1V_1-W_2V_2)\\
-i\bigl(\frac{2}{3}\Lambda_2[\bm{W}_\bot\!\times\!\bm{V}_\bot]_3
+\frac{1}{3}[\bm{\Gamma}_\bot\!\times\!\bm{\Lambda}_\bot]_3\bigl)\biggl]
\biggl\}
\label{c112}
\end{multline}
\begin{multline}
C^{\frac{3}{2}\frac{1}{2}}_{11}(\bm{\lambda}_\bot,\bm{\Lambda}_\bot;
\mu,\mu^\prime)=
\frac{1}{D}\frac{1}{\sqrt{2}}\biggl\{\\
\times\delta_{\mu,\mu^\prime}
\biggl[\frac{1}{3}(m+(1\!-\!\eta){\cal M}_0)\Gamma_1
-i[\bm{\Gamma}_\bot\!\times\!\bm{\Lambda}_\bot]_3\!\biggl]\\
+\delta_{\mu,\mu^\prime}(-1)^{\frac{1}{2}-\mu^\prime}
\biggl[\frac{2}{3}(m+(1\!-\!\eta){\cal M}_0)\bm{W}_\bot\!\cdot\!\bm{V}_\bot\\
+(\Gamma_1\Lambda_1+\frac{1}{3}\Gamma_2\Lambda_2)
-i\frac{1}{3}(m+(1\!-\!\eta){\cal M}_0)\Gamma_2\biggl]\\
+\delta_{\mu,-\mu^\prime}\biggl[\frac{1}{3}(m+(1\!-\!\eta){\cal M}_0)\Gamma_1\\
+\frac{2}{3}\Lambda_1(u-\bm{W}_\bot\!\cdot\!\bm{V}_\bot)
+\frac{2}{3}\Lambda_1(W_1V_1-W_2V_2)\\
+i\bigl(\frac{2}{3}\Lambda_2[\bm{W}_\bot\!\times\!\bm{V}_\bot]_3
+\frac{1}{3}(\Gamma_2\Lambda_1+\Gamma_1\Lambda_2)\bigl)\biggl]\\
+\delta_{\mu,-\mu^\prime}(-1)^{\frac{1}{2}-\mu^\prime}
\biggl[-\frac{2}{3}\Lambda_1[\bm{W}_\bot\!\times\!\bm{V}_\bot]_3\\
+\frac{1}{3}(\Gamma_1\Lambda_1-\Gamma_2\Lambda_2)
+i\bigl(\frac{1}{3}(m+(1\!-\!\eta){\cal M}_0)\Gamma_2\\
+\frac{2}{3}\Lambda_2(u-\bm{W}_\bot\!\cdot\!\bm{V}_\bot)
-\frac{2}{3}\Lambda_2(W_1V_1-W_2V_2)\bigl)\biggl]\\
+\frac{\Delta_-}{\sqrt{3}}\biggl[-2(m+(1\!-\!\eta){\cal M}_0)(W_1V_1-W_2V_2)\\
+\Gamma_1\Lambda_1-\Gamma_2\Lambda_2\biggl]\\
+\frac{i\Delta_+}{\sqrt{3}}\biggl[-2(m+(1\!-\!\eta){\cal M}_0)
[\bm{W}_\bot\!\times\!\bm{V}_\bot]_3\\
+\Gamma_2\Lambda_1+\Gamma_1\Lambda_2\biggl]
+\frac{\delta_-}{\sqrt{3}}\biggl[\bm{\Gamma}_\bot\!\cdot\!\bm{\Lambda}_\bot
+i\Lambda_2u\biggl]\\
+\frac{\delta_+}{\sqrt{3}}\biggl[2\Lambda_1u
-i[\bm{\Gamma}_\bot\!\times\!\bm{\Lambda}_\bot]_3\!\biggl]
\biggl\}
\label{c132}
\end{multline}
where ${\bm\lambda}_\bot$ and $\bm{\Lambda}_\bot$ are defined 
in Eq.~(\ref{lam}) and we use the following notations: 
\begin{equation}
\bm{W}_\bot=\bm{\lambda}_\bot+\xi\bm{\Lambda}_\bot,\quad
\bm{V}_\bot=\bm{\lambda}_\bot-(1-\xi)\bm{\Lambda}_\bot,
\label{wv}
\end{equation}
\begin{equation}
\bm{\Gamma}_\bot=\eta(1-\xi){\cal M}_0\bm{\lambda}_\bot
+(m+(1\!-\!\eta){\cal M}_0)\bm{\Lambda}_\bot,
\label{gg}
\end{equation}
\begin{equation}
u=(m+\eta\xi{\cal M}_0)(m+\eta(1-\xi){\cal M}_0)
\label{u}
\end{equation}
\begin{equation}
\Delta_\pm(\mu,\mu^\prime)=\delta_{\mu,-3/2}\delta_{\mu^\prime,1/2}\pm
\delta_{\mu,3/2}\delta_{\mu^\prime,-1/2}
\label{Dpm}
\end{equation}
\begin{equation}
\delta_\pm(\mu,\mu^\prime)=\delta_{\mu,-3/2}\delta_{\mu^\prime,-1/2}\pm
\delta_{\mu,3/2}\delta_{\mu^\prime,1/2}
\label{dpm}
\end{equation}
\begin{multline}
N_0=m(m+(1-\!\eta){\cal M}_0)+\xi(1-\xi)\eta^2{\cal M}_0^2\\
-(\bm{\lambda}_\bot+\xi\bm{\Lambda}_\bot)\!\cdot\!
(\bm{\lambda}_\bot-(1-\xi)\bm{\Lambda}_\bot),
\label{n0}
\end{multline}
\begin{multline}
N^\prime_0
=m(m+(1-\!\eta){\cal M}_0^\prime)+\xi(1-\xi)\eta^2{{\cal M}_0^\prime}^2\\
-(\bm{\lambda}_\bot+\xi\bm{\Lambda}^\prime_\bot)\!\cdot\!
(\bm{\lambda}_\bot-(1-\xi)\bm{\Lambda}^\prime_\bot),
\label{n0p}
\end{multline}
\begin{equation}
D=\prod_{i=1}^3d_i ,\quad d_i=\sqrt{(m_i+x_i{\cal M}_0)^2+\bm{k}_{\bot i}^2}.
\label{di}
\end{equation}

The matrix $A_{\bar\mu,\mu}$ used in Sect.~\ref{s24} in Eq.~(\ref{jp}) 
has a form
\begin{equation}
A_{\bar\mu,\mu}=
\begin{cases}
-\frac{1}{3}\delta_{\mu,-\bar\mu}(-1)^{1/2-\mu}, \quad \mbox{if}\,\, 
S=\frac{1}{2},& \\
\frac{2}{3}(\delta_{\mu,\frac{1}{2}}\delta_{\bar\mu,-\frac{1}{2}}
-\delta_{\mu,-\frac{1}{2}}\delta_{\bar\mu,\frac{1}{2}})&\\
-\sqrt{\frac{1}{3}}(\delta_{\mu,-\frac{3}{2}}\delta_{\bar\mu,-\frac{1}{2}}
-\delta_{\mu,-\frac{1}{2}}\delta_{\bar\mu,-\frac{3}{2}})&\\ 
+\sqrt{\frac{1}{3}}(\delta_{\mu,\frac{3}{2}}\delta_{\bar\mu,\frac{1}{2}}
-\delta_{\mu,\frac{1}{2}}\delta_{\bar\mu,\frac{3}{2}}), \,\,
\mbox{if}\,\,S=\frac{3}{2}.&
\end{cases}
\label{amumu}
\end{equation}

\newpage

\begin{widetext}

\begin{figure}[hp]
\begin{center}
\epsfig{figure=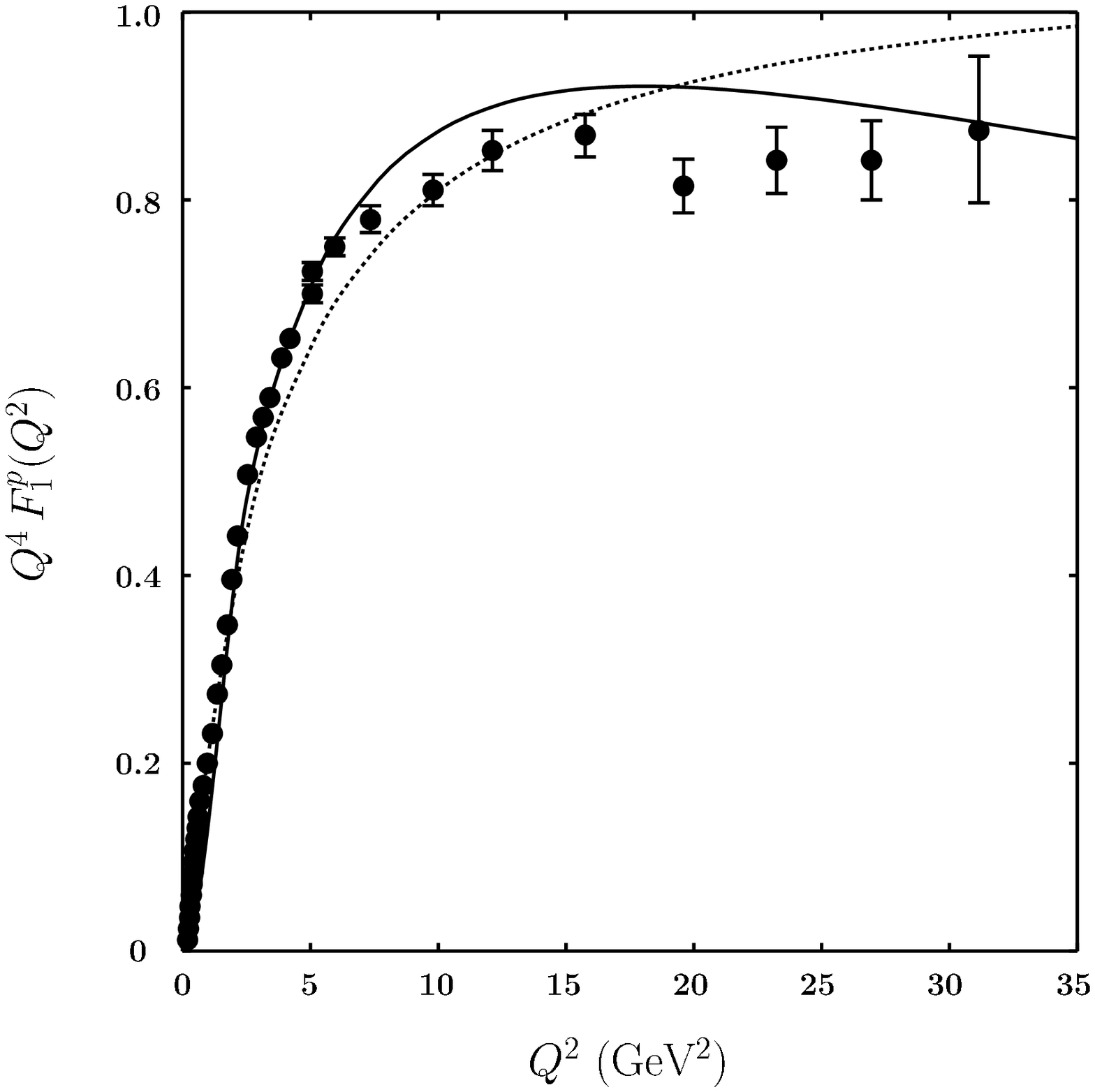,width=0.7\textwidth,clip}
\end{center}
\caption{Proton Dirac form factor multiplied with $Q^4$. Experimental data
are taken from Ref.~\cite{diehl06}. Prediction of the light-front quark model 
is given by the solid line, results from the AdS/QCD 
approach~\cite{Gutsche:2012bp} are marked by the dotted line.} 
\label{f1}
\end{figure}
\begin{figure}[hp]
\begin{center}
\epsfig{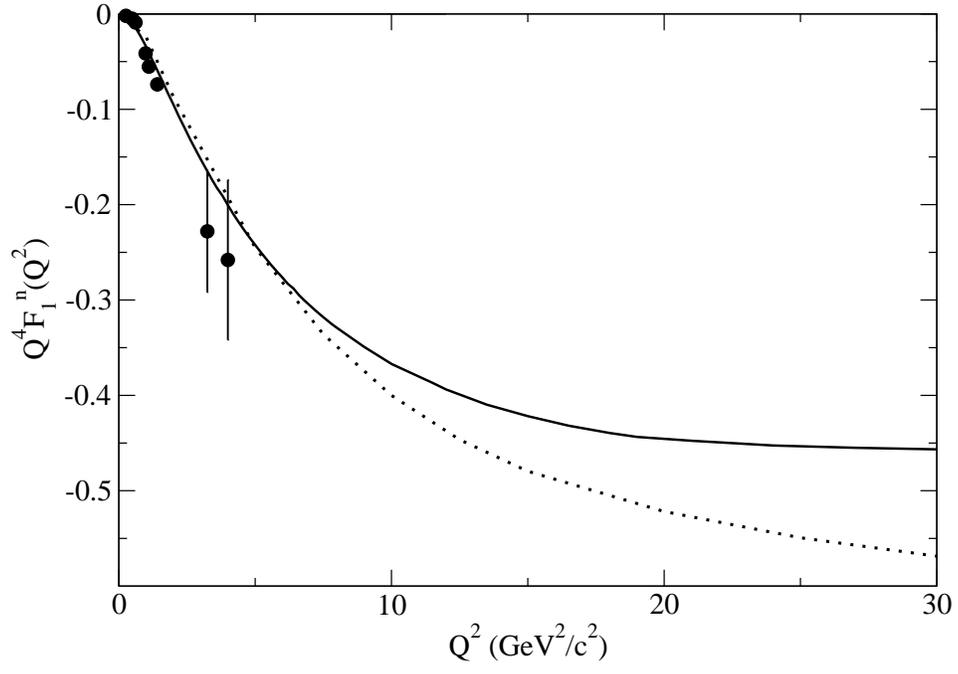}
\end{center}
\caption{Neutron Dirac form factor multiplied with $Q^4$. Experimental data
are taken from Ref.~\cite{diehl06}. Prediction of the light-front quark model 
is given by the solid line, results from the AdS/QCD 
approach~\cite{Gutsche:2012bp} are marked by the dotted line.}  
\label{f1n}
\end{figure}
\begin{figure}[hp]
\begin{center}
\epsfig{figure=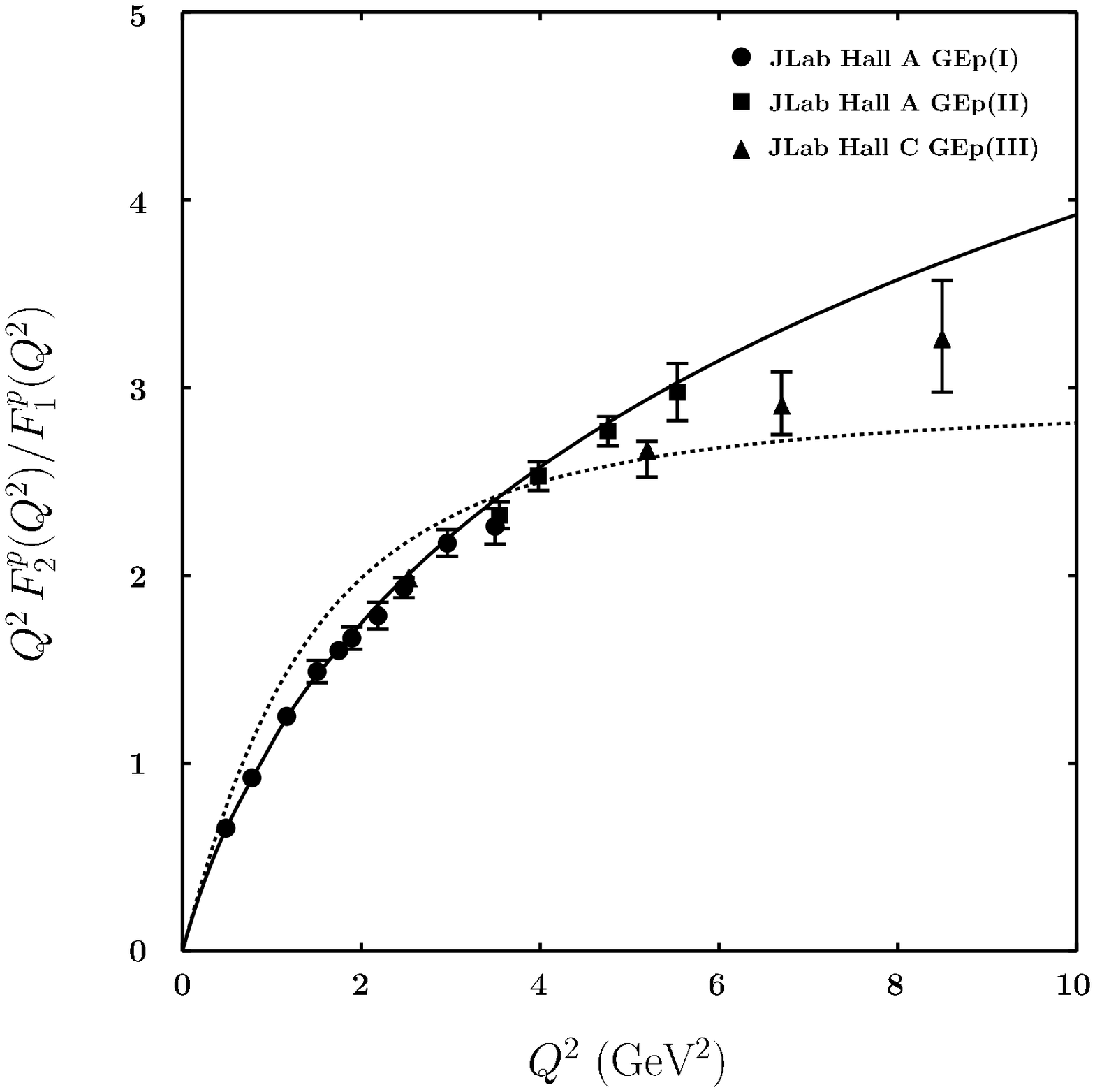,width=0.7\textwidth,clip}
\end{center}
\caption{Results for $Q^2F^p_2(Q^2)/F^p_1(Q^2)$. Experimental data are taken 
from Refs.~\cite{milbrath98,jones00,punjabi05,ron11,crowford07,%
jones06,maclachlan06,jahn11,puckett10}. 
Prediction of the light-front quark model is given
by the solid line, results from the AdS/QCD 
approach~\cite{Gutsche:2012bp} are marked by the dotted line.} 
\label{f3}
\end{figure}
\begin{figure}[hp]
\begin{center}
\epsfig{figure=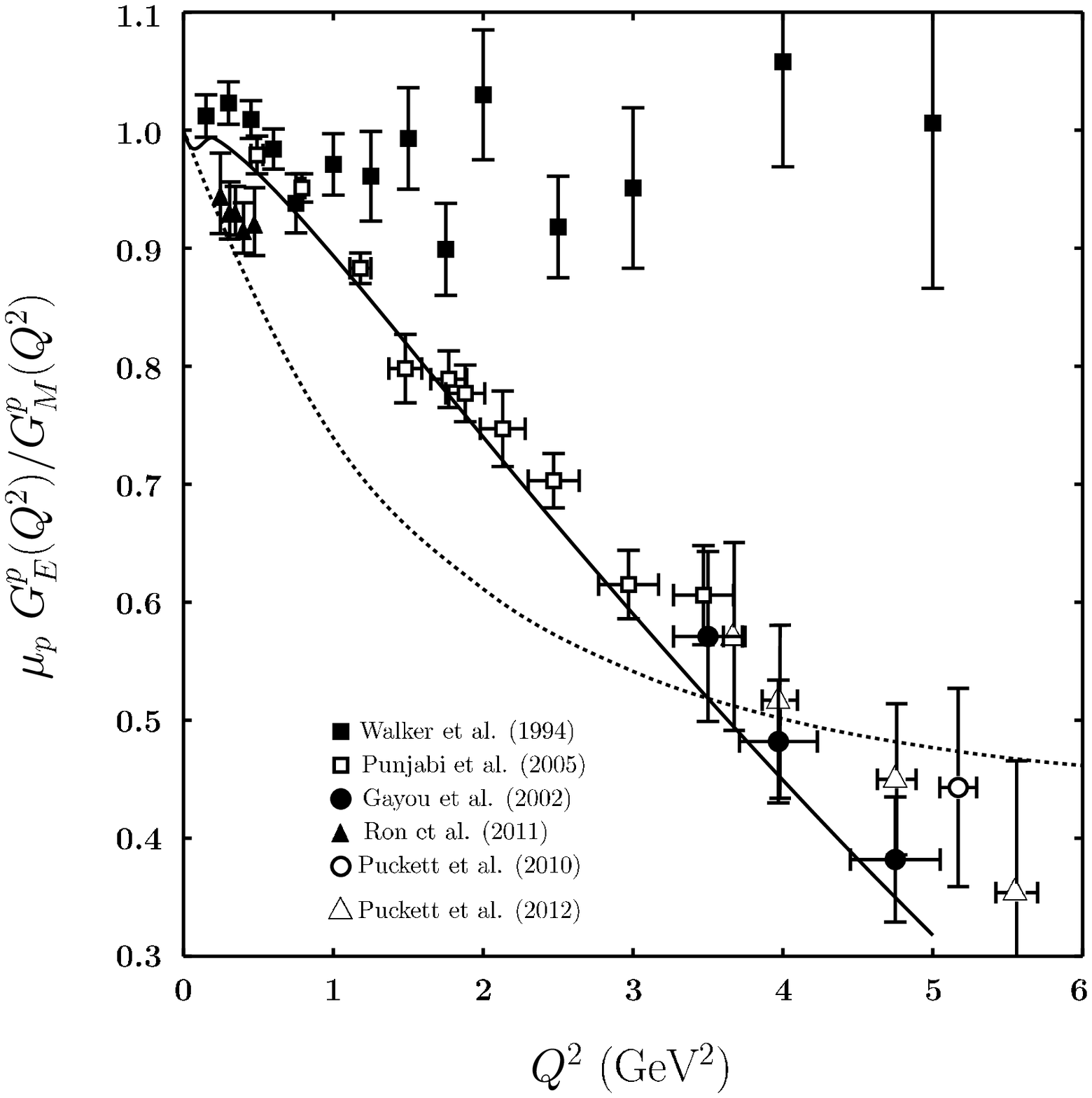,width=0.7\textwidth,clip}
\end{center}
\caption{Ratio $\mu_pG^p_E(Q^2)/G^p_M(Q^2)$ in comparison to the experimental
data taken from Refs.~\cite{gayou02,walker94,punjabi05,puckett12,ron11}. 
Prediction of the light-front quark model is given
by the solid line, results from the AdS/QCD 
approach~\cite{Gutsche:2012bp} are marked by the dotted line.} 
\label{f4}
\end{figure}
\begin{figure}[hp]
\begin{center}
\epsfig{figure=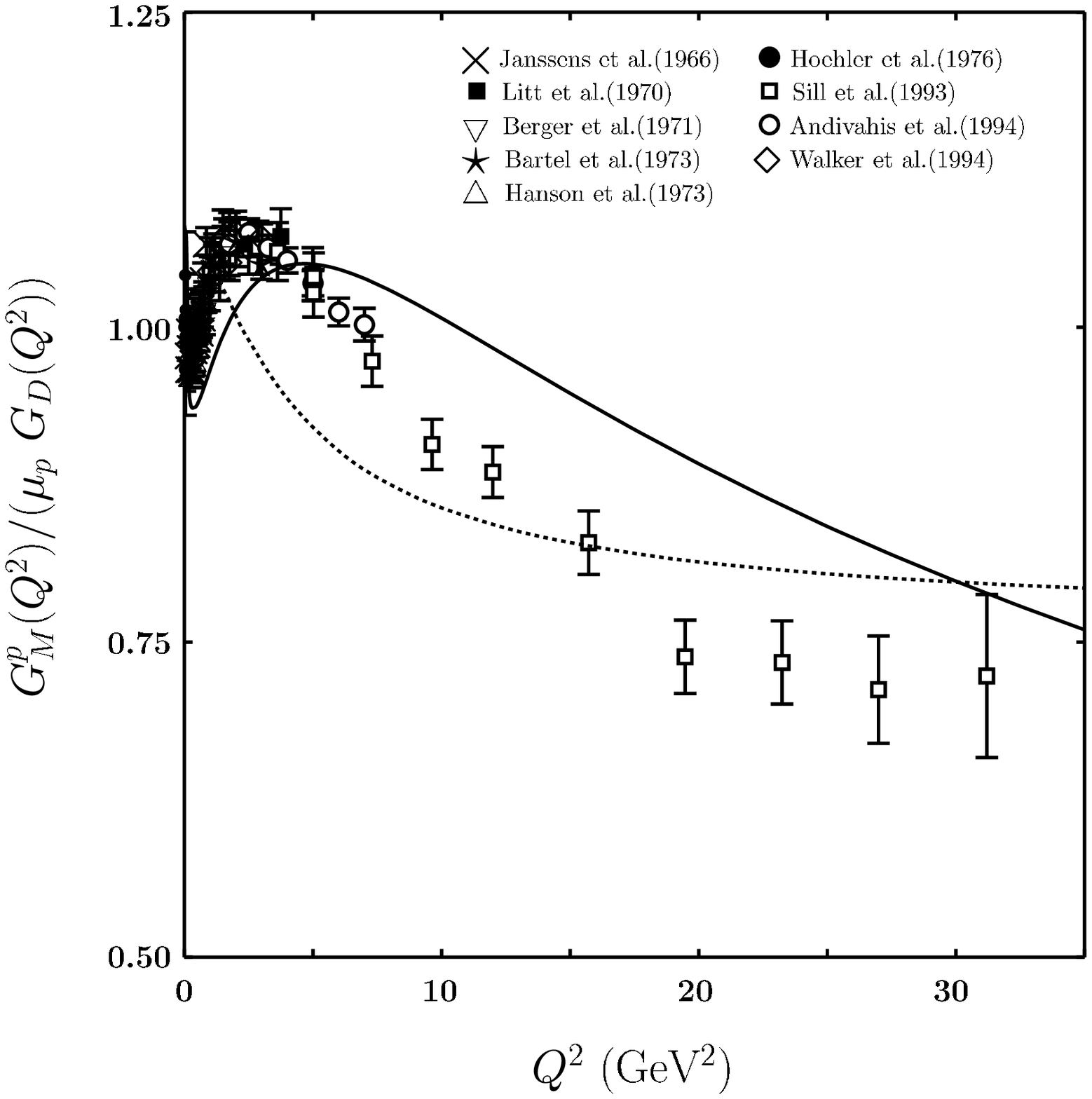,width=0.7\textwidth,clip}
\end{center}
\caption{Ratio $G^p_M(Q^2)/(\mu_pG_D(Q^2))$. Experimental data are taken from
Refs.~\cite{gayou02,walker94,punjabi05}. 
Prediction of the light-front quark model is given
by the solid line, results from the AdS/QCD 
approach~\cite{Gutsche:2012bp} are marked by the dotted line.}
\label{f2}
\end{figure}
\begin{figure}[hp]
\begin{center}
\epsfig{figure=fig6.eps,width=0.7\textwidth,clip}
\end{center}
\caption{Ratio $G^n_M(Q^2)/(\mu_nG_D(Q^2))$. Experimental data are taken from
Refs.~\cite{lachniet09,bartel73,anderson07,arnold88,lung93,kubon02,anklin98}. 
Prediction of the light-front quark model is given
by the solid line, results from the AdS/QCD 
approach~\cite{Gutsche:2012bp} are marked by the dotted line.}
\label{f2n}
\end{figure}
\begin{figure}[hp]
\begin{center}
\epsfig{figure=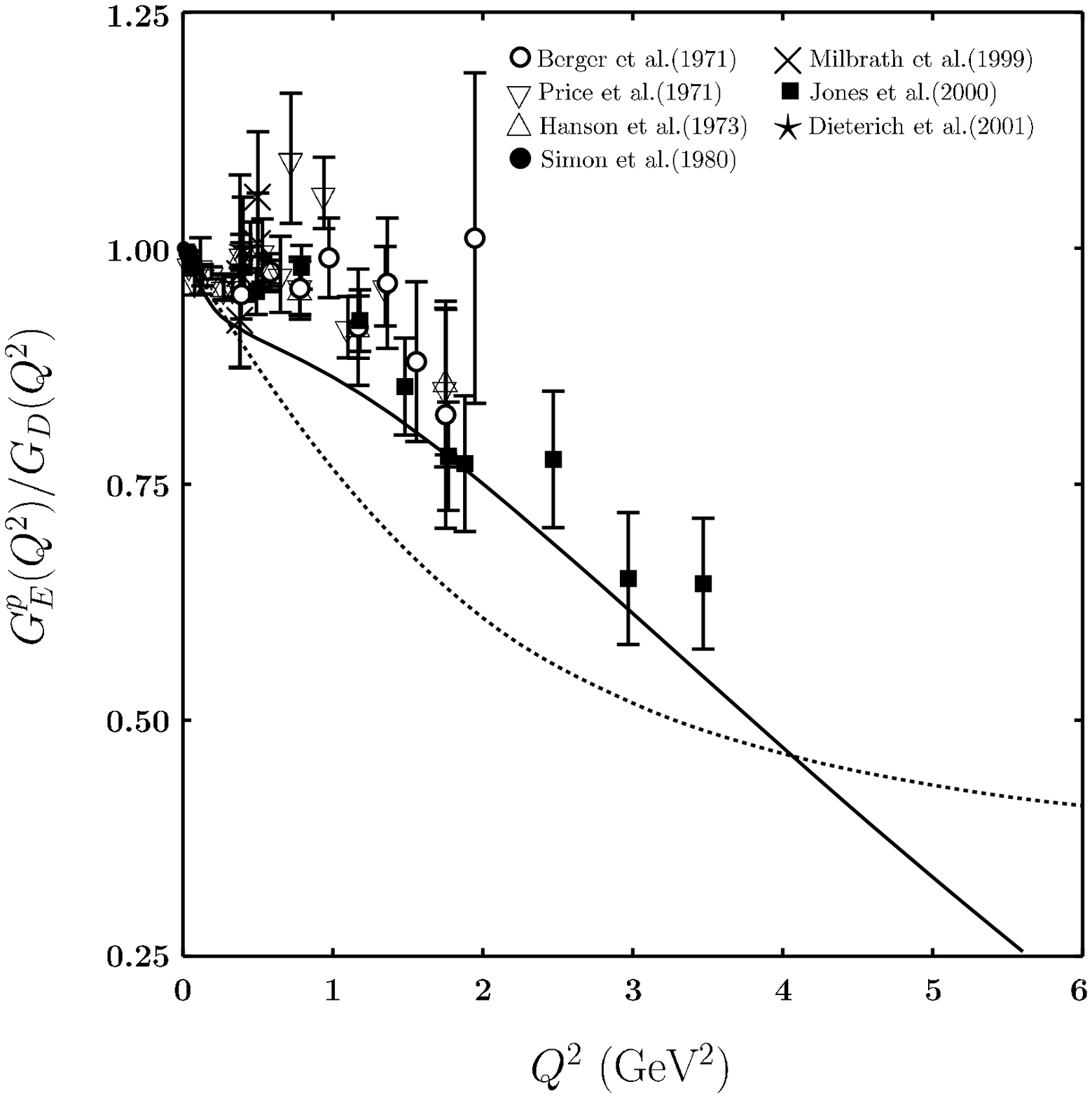,width=0.7\textwidth,clip}
\end{center}
\caption{Ratio $G^p_E(Q^2)/G_D(Q^2)$. Experimental data are taken from 
Refs.~\cite{simon80,price71,berger71,hanson73,milbrath98,dieterich01,%
jones00,gayou02}. 
Prediction of the light-front quark model is given
by the solid line, results from the AdS/QCD 
approach~\cite{Gutsche:2012bp} are marked by the dotted line.} 
\label{f5}
\end{figure}
\begin{figure}[hp]
\begin{center}
\epsfig{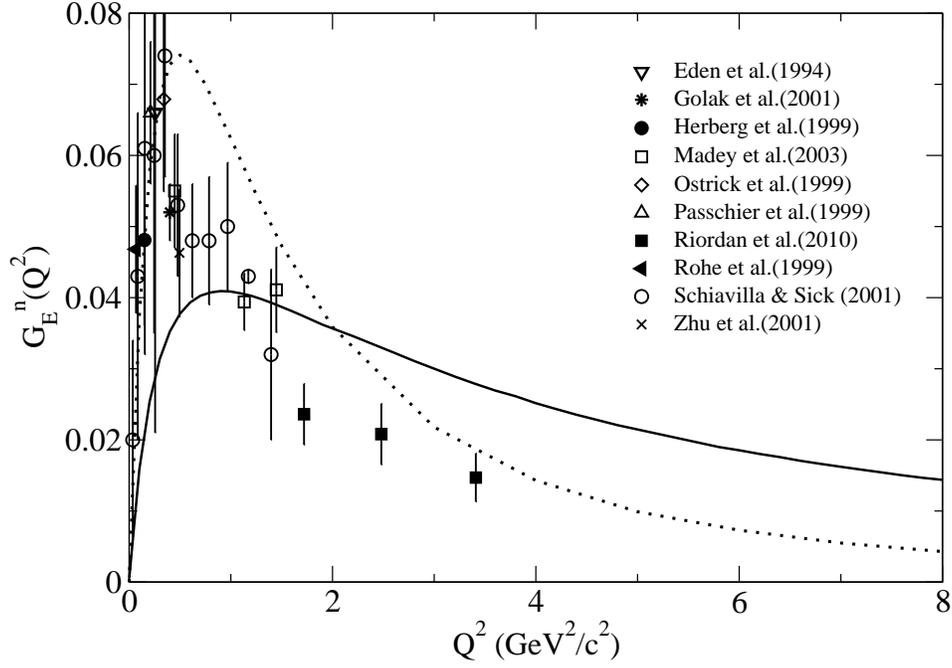}
\end{center}
\caption{The charge neutron form factor $G^n_E(Q^2)$. Experimental data are
taken from Refs.~\cite{herberg99,passchier99,eden94,ostrick99,golak01,%
madey03,zhu01,rohe99,schiavilla01}. 
Prediction of the light-front quark model is given
by the solid line, results from the AdS/QCD 
approach~\cite{Gutsche:2012bp} are marked by the dotted line.} 
\label{f5n}
\end{figure}
\begin{figure}
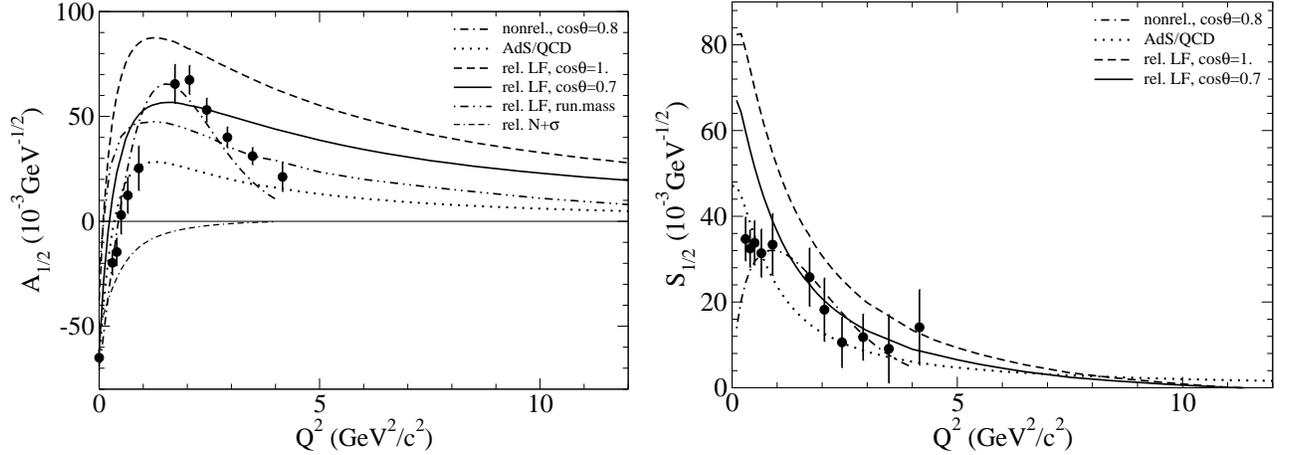

\begin{center}
\mbox{
{\epsfig{figure=fig9.eps,width=0.46\textwidth,clip}}\quad
{\epsfig{figure=fig10.eps,width=0.46\textwidth,clip}}
}
\caption{The helicity amplitudes $A_{1/2}$ and $S_{1/2}$ 
(on the left and right 
pannels, respectively) for electroproduction of the Roper resonance on the 
proton. The data are from~\cite{aznauryan09,mokeev09}. 
The predictions of the light-front quark model are given by the short-dashed
($\cos\theta =$1),  and the solid lines ($\cos\theta=$0.7); 
results from the AdS/QCD 
approach~\cite{Gutsche:2012bp} are marked by the dotted line. Nonrelativistic 
results of Ref.~\cite{obukhovsky11} (the dashed dotted line) and the result of
Ref.~\cite{aznauryan12} for LF model with running quark masses (the 
double-dotted dashed line) are also shown for comparison. The $A_{1/2}$ 
amplitude for hadronic $N\!+\!\sigma$ molecular state are marked by the 
double-dashed dotted line (adopted from ref.~\cite{obukhovsky11}).} 
\label{fr12}
\end{center}
\end{figure}

\end{widetext}

\end{document}